\documentclass{article} 
\usepackage{cite} 
\usepackage{iclr2023_conference,times}
\usepackage{graphicx}
\usepackage{array}
\usepackage{booktabs}
\usepackage{multirow}
\usepackage{colortbl}
\usepackage{soul}
\usepackage{makecell}
\usepackage{xcolor}
\usepackage{xspace}
\usepackage{array}
\usepackage{adjustbox}
\usepackage{tabularx}
\usepackage{subcaption}
\usepackage{tikz}

\usepackage{amssymb}
\usepackage{pifont}
\newcommand{\cmark}{\ding{51}}%
\newcommand{\xmark}{\ding{55}}%

\usepackage{amsmath,amsfonts,bm}

\def\eqref#1{equation~\ref{#1}}

\def\1{\bm{1}}

\DeclareMathAlphabet{\mathsfit}{\encodingdefault}{\sfdefault}{m}{sl}
\SetMathAlphabet{\mathsfit}{bold}{\encodingdefault}{\sfdefault}{bx}{n}

\usepackage{hyperref}
\usepackage{url}
\usepackage{booktabs,caption}
\usepackage[flushleft]{threeparttable}

\newcommand{\ourmodel}{\textit{MedImageInsight}\xspace}
\newcommand{\ourmodelshort}{\textit{MI2}\xspace}
\title{\ourmodel: An Open-Source Embedding Model for General Domain Medical Imaging}

\author{Noel C. F. Codella\textsuperscript{1}, Ying Jin\textsuperscript{1}, Shrey Jain\textsuperscript{1}, Yu Gu\textsuperscript{1}, Ho Hin Lee\textsuperscript{1}, Asma Ben Abacha\textsuperscript{1},  \\ \textbf{Alberto Santamaria-Pang\textsuperscript{1},  Will Guyman\textsuperscript{1}, Naiteek Sangani\textsuperscript{1}, Sheng Zhang\textsuperscript{2}, Hoifung Poon\textsuperscript{2}, }\\  \textbf{  Stephanie Hyland\textsuperscript{2}, Shruthi Bannur\textsuperscript{2}, Javier Alvarez-Valle\textsuperscript{2}, Xue Li\textsuperscript{4}, John Garrett\textsuperscript{4},}\\ \textbf{  Alan McMillan\textsuperscript{4}, Gaurav Rajguru\textsuperscript{3}, Madhu Maddi\textsuperscript{3}, Nilesh Vijayrania\textsuperscript{3}, Rehaan Bhimai\textsuperscript{3},}\\
\textbf{  Nick Mecklenburg\textsuperscript{3}, Rupal Jain\textsuperscript{3}, Daniel Holstein\textsuperscript{3}, Naveen Gaur\textsuperscript{3}, Vijay Aski\textsuperscript{3},} \\
\textbf{ Jenq-Neng Hwang\textsuperscript{5}, Thomas Lin\textsuperscript{1}, Ivan Tarapov\textsuperscript{1}, Matthew Lungren\textsuperscript{1,*}, Mu Wei\textsuperscript{1,*}} \\
\\
\textsuperscript{1}Microsoft Health and Life Sciences, Redmond, WA USA \\
\textsuperscript{2}Microsoft Research, Redmond, WA USA \\
\textsuperscript{3}Microsoft Azure AI, Redmond, WA USA \\
\textsuperscript{4}University of Wisconsin-Madison, Madison, WI \\
\textsuperscript{5}University of Washington, Seattle, WA  \\
\textsuperscript{*}Equal Contribution \\
\\
}

\iclrfinalcopy 

\begin{document}
\maketitle

\begin{abstract}

In this work, we present \ourmodel, an open-source medical imaging embedding model. \ourmodel is trained on medical images with associated text and labels across a diverse collection of domains, including X-Ray, CT, MRI, dermoscopy, OCT, fundus photography, ultrasound, histopathology, and mammography. Rigorous evaluations demonstrate \ourmodel's ability to achieve state-of-the-art (SOTA) or human expert level performance across classification, image-image search, and fine-tuning tasks.  Specifically, on public datasets, \ourmodel achieves SOTA in CT 3D medical image retrieval, as well as SOTA in disease classification and search for chest X-ray, dermatology, and OCT imaging. Furthermore, \ourmodel achieves human expert performance in bone age estimation (on both public and partner data), as well as AUC above 0.9 in most other domains. When paired with a text decoder, \ourmodel achieves near SOTA level single image report findings generation with less than 10\% the parameters of other models. Compared to fine-tuning GPT-4o with only MIMIC-CXR data for the same task, \ourmodel outperforms in clinical metrics, but underperforms on lexical metrics where GPT-4o sets a new SOTA. Importantly for regulatory purposes, \ourmodel can generate ROC curves, adjust sensitivity and specificity based on clinical need, and provide evidence-based decision support through image-image search (which can also enable retrieval augmented generation). In an independent clinical evaluation of image-image search in chest X-ray, \ourmodel outperformed every other publicly available foundation model evaluated by large margins (over 6 points AUC), and significantly outperformed other models in terms of AI fairness (across age and gender). We hope releasing \ourmodel will help enhance collective progress in medical imaging AI research and development.

\newpage

\end{abstract}

\section{Introduction}
\label{introduction}

Over the past two decades, the generation of medical image data has surged at an annual growth rate of 6\%, significantly outpacing the 0.7\% annual growth rate of the medical imaging workforce ~\cite{growthstats}. This disparity underscores a critical challenge in the healthcare industry: the rapid increase in imaging data together with the slower workforce growth highlights the need for enhanced efficiency and productivity. Integrating AI and machine learning into imaging workflows can address these challenges by improving diagnostic accuracy, automating routine tasks, and supporting clinical decision-making~\cite{humanai}. The recent advancements in foundation models ~\cite{gpt3,gpt4,gpt4o,gemini} have sparked considerable interest in their application to healthcare. Foundation models offer the advantage of generalizing across multiple sub-domains and tasks, thereby potentially reducing development costs for a wide range of clinical scenarios. Historically, AI applications in healthcare have been limited by the inability of models to scale across numerous tasks and sub-domains, necessitating the development of individual models for each task. 

Preliminary investigations indicate that foundation models can scale rapidly, as evidenced by GPT-4’s performance on the USMLE and other exams, encompassing a diverse array of healthcare sub-domains ~\cite{msrgpt4,usmle,medprompt}. Simultaneously, numerous studies have demonstrated the progressive advancements of foundation models in image-text medical imaging datasets and benchmarks ~\cite{PMC15M,llava-med,llavarad,maira1,maira2,cxrfoundation,elixr,medgemini,medpalm}. Models like GLoRIA ~\cite{gloria}, CXR Foundation ~\cite{cxrfoundation}, and ELIXR ~\cite{elixr} were trained on chest X-ray datasets using contrastive and generative objective functions, enabling effective fine-tuning as well as inherent generative and classification capabilities within that domain. BiomedCLIP~\cite{PMC15M} employed a contrastive pretraining approach on a corpus of medical image-caption pairs from PubMed, establishing robust performance for downstream tasks in histopathology, chest X-ray pneumonia detection, and VQA in X-ray, CT, and MRI. LLaVA-Med~\cite{llava-med}, an encoder-decoder architecture akin to LLaVA, was further trained on diverse medical imaging data, including X-ray, CT, MRI, histology, and gross pathology, using GPT-4 to generate dialogue for instruction-tuning to support chat-based interactions. Med-Gemini~\cite{medgemini} and Med-PaLM-M~\cite{medpalm} are multimodal LLMs designed to unify multiple text-based and multimodal medical tasks under a single model. BiomedGPT~\cite{biomedgpt} is a recent encoder-decoder structure that is pre-trained using several objectives, including masked-image-modeling, image-infilling, object detection, VQA, and captioning. While BiomedGPT has diverse inherent image-text capability across many medical imaging domains, for classification tasks (with support for generating ROC curves) it requires fine-tuning. LLaVA-Rad ~\cite{llavarad} was similar to LLaVA-Med, but focused on chest X-ray report generation. MAIRA-1~\cite{maira1} and MAIRA-2~\cite{maira2} are other encoder-decoder generative architectures that have advanced the state-of-the-art in chest X-ray report generation. In MAIRA-2, the model incorporates image-text grounding and comparisons to prior imaging studies.

\begin{table}[t!]
  \centering
  \begin{threeparttable}
    \caption{Medical Imaging Foundation Model Comparisons\textsuperscript{*}}
    \label{tab:comp}
    \begin{tabular}{@{}llccccccc@{}}
      \toprule
      Model & Params & \begin{tabular}{@{}c@{}}Multi-\\Domain\textsuperscript{3}\end{tabular} & \begin{tabular}{@{}c@{}}Text\\Pretrain\end{tabular} & \begin{tabular}{@{}c@{}}Label\\Pretrain\end{tabular} & ROC\textsuperscript{2} & I2I\textsuperscript{2} & Gen. & 3D \\
      \midrule 
      CXR Foundation ~\cite{cxrfoundation} & 0.25 B & \xmark  & \xmark & {\color{green} \cmark} & {\color{green} \cmark} & \xmark & \xmark & \xmark  \\
      ELIXR ~\cite{elixr} & 0.25 B & \xmark & {\color{green} \cmark} & {\color{green} \cmark} & {\color{green} \cmark} & \xmark  & {\color{green} \cmark} & \xmark \\
      BiomedCLIP ~\cite{PMC15M} & 0.09 B & {\color{green} \cmark}& {\color{green} \cmark} & \xmark & {\color{green} \cmark} & \xmark  & {\color{green} \cmark} & \xmark \\
      MAIRA-1 ~\cite{maira1} & 7 B & \xmark  & {\color{green} \cmark} & \xmark & \xmark  & \xmark   & {\color{green} \cmark} & \xmark \\
      MAIRA-2 ~\cite{maira2} & 7-13 B & \xmark & {\color{green} \cmark} & \xmark & \xmark  & \xmark   & {\color{green} \cmark} & \xmark \\
      LLaVA-Med ~\cite{llava-med} & 7 B & {\color{green} \cmark}& {\color{green} \cmark} & \xmark & \xmark  & \xmark   & {\color{green} \cmark} & \xmark \\
      LLaVA-Rad ~\cite{llavarad} &  7 B & \xmark & {\color{green} \cmark} & \xmark & \xmark  & \xmark   & {\color{green} \cmark} & \xmark \\
      Med-Gemini ~\cite{medgemini} & NR\textsuperscript{1} & {\color{green} \cmark} & {\color{green} \cmark} & {\color{green} \cmark} & \xmark  & \xmark   & {\color{green} \cmark} & \xmark \\
      Med-PaLM-M ~\cite{medpalm} & 84-562 B & {\color{green} \cmark}&{\color{green} \cmark} & {\color{green} \cmark} & \xmark  & \xmark   & {\color{green} \cmark} & \xmark \\
      BiomedGPT ~\cite{biomedgpt} & 0.18 B & {\color{green} \cmark}&{\color{green} \cmark} & \xmark & {\color{green} \cmark}  & \xmark   & {\color{green} \cmark} & \xmark \\

    \midrule

    \ourmodel  & $<$0.61 B\textsuperscript{4} & {\color{green} \cmark} & {\color{green} \cmark} & {\color{green} \cmark} & {\color{green} \cmark} & {\color{green} \cmark}  & {\color{green} \cmark} & {\color{green} \cmark}\\

      \bottomrule
    \end{tabular}
    \begin{tablenotes}
      \small
      \item[*] ROC = Ability to generate ROC curves for regulatory compliance and sensitivity/specificity adjustments. I2I = Optimized and evaluated for image-image search capability. Gen. = Generative capable. 3D = Trained and evaluated on 3D medical image retrieval task.  NR = Not reported.
      \item[1] Med-Gemini model sizes have not been reported, but are expected to be many billions given size of the smallest version, Gemini Nano (3.25 B).
      \item[2] The ability of models to support generation of ROC curves and provide evidence based decisions through image-image search are two critical features to comply with current regulatory requirements and guidance on model transparency~\cite{whitehouse2023ai}. 
      \item[3] Model is trained across multiple medical imaging domains (i.e. Chest X-Ray, Abdominal CT, Dermatology, Pathology, etc.). 
      \item[4] Image-image search relies on only the 0.36B image encoder. Image classification relies on the 0.36B image encoder for each query image and the 0.25B text encoder for each text prompt (one time computation for all images). Report generation relies on 0.36B image encoder and 0.07B text decoder. See Fig. ~\ref{fig:arch}.
    \end{tablenotes}
  \end{threeparttable}
\end{table}

Despite significant advancements in applying foundation models to healthcare, several challenges persist. Firstly, generative foundation models, while offering flexibility in supporting various interaction modes and styles, inherently lack widely adopted methods to associate classification decisions with confidence scores. These scores are essential for generating ROC curves for evaluation, meeting regulatory compliance, and customizing sensitivity/specificity tradeoffs ~\cite{whitehouse2023ai}. Secondly, current models designed for classification and text generation do not inherently provide direct evidence to support predictions or offer genuine decision transparency (beyond another generative output), which is crucial for clinical workflow integration ~\cite{pranavhumani}. Lastly, existing works do not study classification, generative, and 2D/3D image retrieval functions from a single encoder model.

\begin{figure}[t!]
    \centering
    \includegraphics[width=0.98\textwidth]{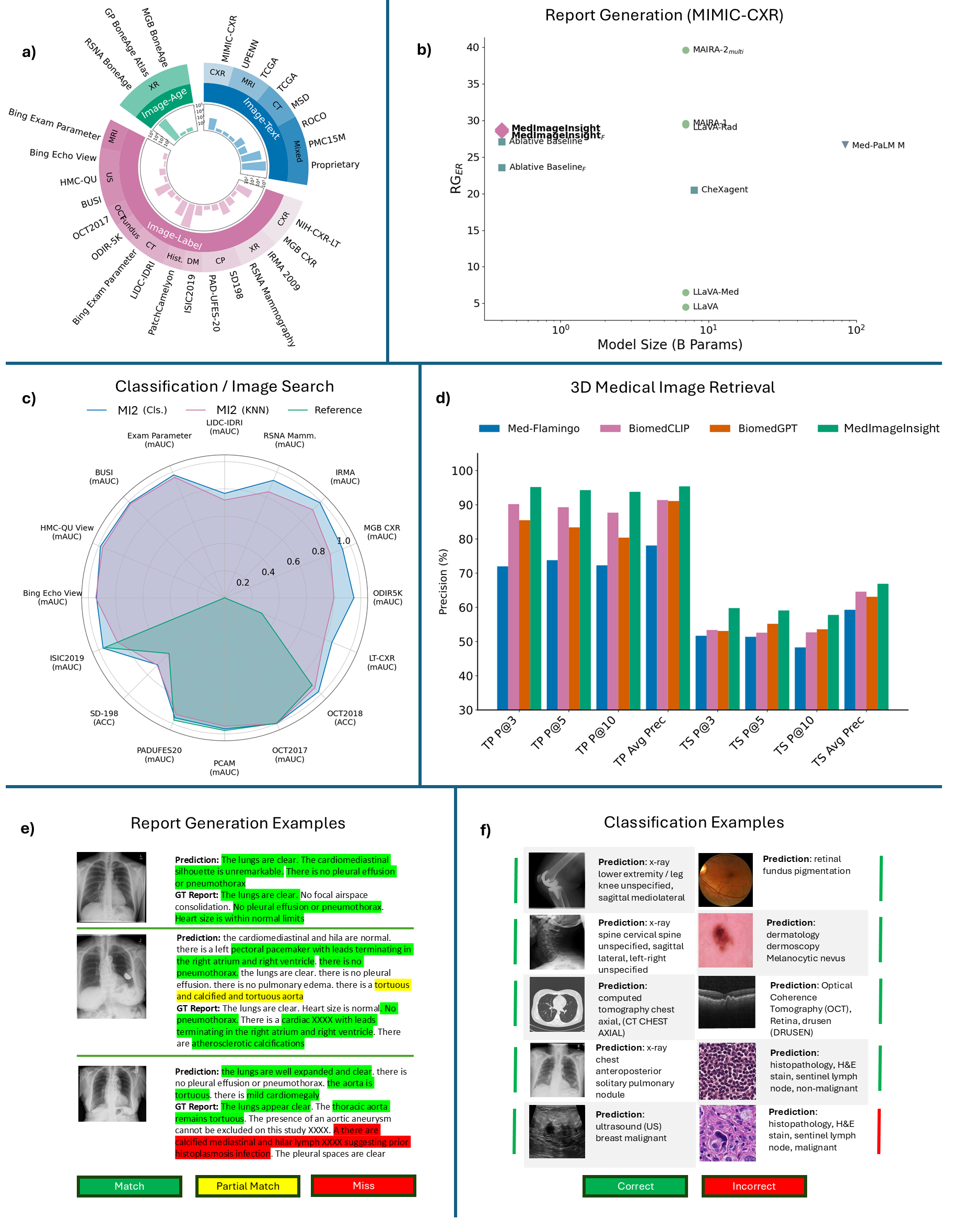}
    \caption{Overview of the \ourmodel foundation model. a) Chord diagram of datasets and modalities used for training and evaluation. b) \ourmodel report findings generation performance among methods that leverage a single image on MIMIC-CXR (except multi-image benchmarks labeled with subscript ``multi''). c) Radar figure of classification performance of single model (no fine-tuning) across datasets. All metrics are mAUC except for SD-198 and OCT2018 which are displayed as accuracy. Reference refers to SOTA. d) 3D retrieval results for 3D-MIR benchmark (inherent capability, no fine-tuning). TP = Tumor Presence. TS = Tumor Stage. P@N = Precision @ N. e) 3 examples of report generation predictions for chest X-ray. f) 10 examples of broad domain medical image classification from roughly 1000 classes (inherent capability, no fine-tuning).}
    \label{fig:overview}
\end{figure}

In this paper we introduce \ourmodel, an open-source generalist model that scales trivially across medical imaging sub-domains while addressing the above mentioned remaining challenges. A visual overview of the datasets and performance of the model are displayed in Fig. ~\ref{fig:overview}.  In summary, the benefits of \ourmodel compared to other works are as follows (and also summarized in Table ~\ref{tab:comp}):

\begin{itemize}
    \item {\bf Efficient Medical Imaging Generalist:} A single lightweight model (0.36B/0.25B parameters image/text encoder) trivially scales to 14 medical imaging domains (without fine-tuning for each domain), in some cases trained with as few as 100 images, commonly reaching SOTA or human-expert level performance on public and partner datasets. Pre-training supports annotation in both the form of image-text pairs or image-label pairs, the two most common forms of annotation available in medical images. 
    \item {\bf Retains ROC Capability:} Classification decisions are  associated with confidence scores that can be used to generate receiver-operating-characteristic (ROC) curves for regulatory approval, or to enable users to adjust sensitivity or specificity levels according to varying clinical needs.
    \item {\bf Inherent Transparency and Evidence:} \ourmodel is trained to support image-image search, which can be used to render classification predictions with "k-Nearest Neighbor" (KNN) algorithms, an inherently explainable and transparent process. KNN is also the foundation of Retrieval Augmented Generation (RAG) approaches for Large Language Models (LLMs). KNN still provides confidence scores and the ability to generate ROC curves.
    \item {\bf Generative-Ready: } The model can be plugged into common decoder structures to provide near SOTA results for single image report findings generation at a fraction of the compute cost of other published models. We demonstrate this using a very lightweight 0.07B parameter decoder, {\em which is 1/100\textsuperscript{th} the size of other studied decoders, keeping the total generative model size below 0.5B parameters}.
    \item {\bf 3D Pre-training and Capability:} We leverage a novel 3D image-text pre-training dataset containing CT images from four key organs of the abdomen and chest (liver, colon, pancreas, and lung), paired with detailed textual descriptions of modality, body region (chest/abdomen), a list of visible organs, as well as the presence or absence of metastatic disease. We demonstrate that \ourmodel, trained with this data, achieves SOTA performance for 3D medical image retrieval.  
\end{itemize}

We believe that by releasing \ourmodel to the open-source community, we can significantly enhance the collective progress in medical imaging AI research and development. This collaborative approach will not only accelerate advancements in medical imaging but also ensure that the benefits of this technology are accessible to a wider audience, as well as improve AI equity and transparency, ultimately leading to greater societal impact and improved healthcare outcomes.

\section{Results}

\subsection{Base Image Search and Text Search}

\definecolor{green25}{rgb}{0.75, 1, 0.75}
\definecolor{yellow25}{rgb}{1, 1, 0.75}
\definecolor{blue25}{rgb}{0.7, 0.9, 1}
\sethlcolor{green25}
\newcommand{\hlgreen}[1]{{\sethlcolor{green25}\hl{#1}}}
\newcommand{\hlyellow}[1]{{\sethlcolor{yellow25}\hl{#1}}}
\newcommand{\hlblue}[1]{{\sethlcolor{blue25}\hl{#1}}}
\begin{table}[h!]
\centering
\small
\begin{threeparttable}
\begin{tabular}{p{1.5cm}p{1.5cm}lp{1.5cm}p{1.0cm}p{1.0cm}p{1cm}}
\toprule
\textbf{Modality} & \textbf{Sub-Modality\textsuperscript{*}} & \textbf{Dataset (\# Labels)} & \textbf{Metric} & {\bf \ourmodelshort (Cls.)**} & {\bf \ourmodelshort (KNN)**} & Ref\textsuperscript{ a} \\ 
\midrule
\multirow{8}{*}{\parbox{1.4cm}{Rad.}} 
& XR & LT-CXR (20) & mAUC & 0.853 & 0.791 &\\
         & &  & BACC & \cellcolor{yellow25} 0.422 & \cellcolor{yellow25} 0.334 & 0.294~\cite{ltcxr}\\
& XR & MGB CXR (80) & mAUC & \cellcolor{blue25} 0.937 & 0.839 &\\
 & XR & IRMA (137) & mAUC & \cellcolor{blue25} 0.989  & \cellcolor{blue25} 0.915  &\\
 & XR & RSNA BoneAge (-)& Ab. L1\textsuperscript{b} & - & \cellcolor{green25} 6.194 & 7.32~\cite{rsnaboneageexpert}\\
  & XR & MGB BoneAge (-)& Ab. L1\textsuperscript{b} & - & \cellcolor{green25} 6.570 & 7.32~\cite{rsnaboneageexpert}\\
 & XR & RSNA Mamm. (4) & mAUC & \cellcolor{blue25} 0.934  & 0.843 &\\
 & CT & LIDC-IDRI (4) & mAUC &  0.767 & 0.718 &\\
 & MR/CT & Exam Parameter (21) &mAUC & \cellcolor{blue25} 0.977  & \cellcolor{blue25} 0.958 &\\
 & US & BUSI (3) & mAUC & \cellcolor{blue25} 0.985 & \cellcolor{blue25} 0.977 &\\
 & US & HMC-QU View (2) & mAUC& \cellcolor{blue25} 0.987 & \cellcolor{blue25} 0.970 &\\
 & US & Bing Echo View (7) & mAUC& \cellcolor{blue25} 0.941 & \cellcolor{blue25} 0.946 &\\
\midrule
\multirow{3}{*}{Derm.} 
 & DM & ISIC2019 (8) & mAUC& \cellcolor{yellow25} 0.966 \textsuperscript{c}  & 0.852 & 0.954~\cite{isic2019sota}\\
 & CP & SD-198 (198) &mAUC & \cellcolor{blue25} 0.989 &  \cellcolor{blue25} 0.952 &\\
 &  &  & ACC & \cellcolor{yellow25} 0.696 \textsuperscript{c} & \cellcolor{yellow25} 0.700 \textsuperscript{c}  & 0.578~\cite{sd198sota}\\
 & CP & PADUFES20 (6) &mAUC & \cellcolor{blue25} 0.954  & \cellcolor{blue25} 0.929  & 0.973\textsuperscript{d}~\cite{medpalm}\\
\midrule
Path. & H\&E & PCAM (2) & mAUC& \cellcolor{blue25} 0.963 & \cellcolor{blue25} 0.943 & 0.975~\cite{pcamsota}\\
\midrule
\multirow{5}{*}{Opth.} & OCT & OCT2017 (4) & mAUC & \cellcolor{yellow25} 1.000 & \cellcolor{yellow25} 1.000 & 0.998~\cite{oct2017sota}\\
& & & ACC & \cellcolor{green25} 1.000 & \cellcolor{green25} 1.000 & 0.952~\cite{oct2017human}\\
& OCT & OCT2018 (4) & mAUC & \cellcolor{yellow25} 0.999 & \cellcolor{yellow25} 0.99 & \\
& & & ACC & \cellcolor{yellow25} 0.975 & \cellcolor{yellow25} 0.936 & 0.909~\cite{biomedgpt}\\
& Fundus & ODIR5K (28) &mAUC & \cellcolor{blue25} 0.949 & 0.803 &\\
\bottomrule
\end{tabular}

\caption{Comprehensive evaluation of \ourmodel (\ourmodelshort) across various datasets, grouped by modality (radiology, dermatology, pathology, ophthalmology) and sub-modality. We report mAUC for all datasets except BoneAge, where it is not applicable (boneage is a retrieval enabled regression problem). Additional metrics are reported according to the metrics specified by each dataset for ranking.  \\ {\em Colormap:} \hlgreen{Exceeds human expert}, \hlyellow{SOTA equivalent or better}, \hlblue{AUC $>$ 0.9}}
\label{table:comprehensive_results}
\begin{tablenotes}
\small
\item[*]   XR = X-Ray, MR = Magnetic Resonance Imaging, CT = Computed Tomography, US = Ultrasound, DM = Dermoscopy, CP = Clinical Photography, OCT = Optical Coherence Tomography
\item[**] Cls. = Classification with a text prompt generated classification head, KNN = K-nearest neighbor image-image search classification and regression. 
\item[a] Reference (``Ref'') refers to SOTA, except in the case of BoneAge, where it refers to human expert performance.
\item[b] BoneAge is reported as average absolute L1 error in months. Smaller error implies better performance.
\item[c] SOTA determination made from recreating same internal test set selection criteria, though exact splits may vary.
\item[d] Med-PaLM-M model uses both image as well as 14 manually measured clinical criteria, some of which may not be available to collect in practice. The model is 84B parameters compared to 0.36B for \ourmodelshort.
\end{tablenotes}
\end{threeparttable}
\end{table}

\begin{table}[h!]
  \centering
  \begin{threeparttable}
    \caption{\ourmodel AI Fairness Analysis}
    \label{tab:rai}
    \begin{tabular}{@{}lccc@{}}
      \toprule
      Dataset & \# Gender\textsuperscript{*} Partition & Metric & Performance \\
      \midrule 
      LTCXR & Female & mAUC & 0.858 \\
       & Male & mAUC & 0.851 \\
      RSNA BoneAge & Female & Ab. L1 & 6.345 \\
       & Male & Ab. L1 & 6.044 \\

      \bottomrule
    \end{tabular}
    \begin{tablenotes}
      \small
      \item[*] As defined by the dataset annotations. 
    \end{tablenotes}
  \end{threeparttable}
\end{table}

Overall model performance across all studied datasets is shown in Table ~\ref{table:comprehensive_results}. Modalities covered include radiology, dermatology, histopathology, optical coherence tomography (OCT), and retinal fundus imaging. 

On public datasets, \ourmodel achieves SOTA or SOTA equivalent results in 5 tasks where there is tracking of competing methods, achieves human expert equivalence or greater in 2 tasks where comparisons to expert performance is available, and achieves mAUC $>$ 0.9 in most of the remaining tasks. 

\begin{figure}[t!]
    \centering
    \includegraphics[width=0.7\textwidth]{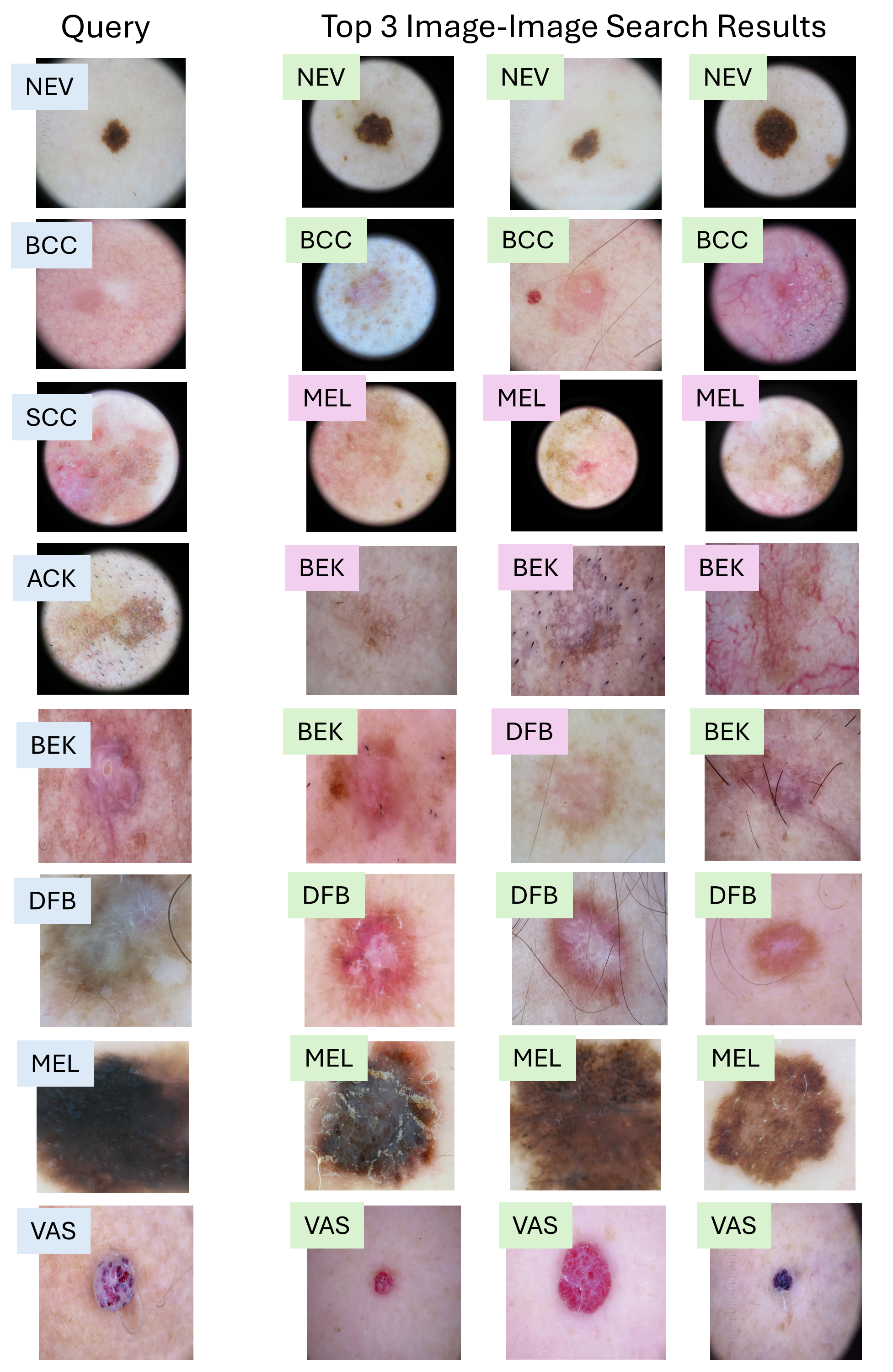}
    \caption{Example image-image search results for the ISIC2019 dataset. 1 query per disease label was randomly selected from the validation dataset. Correctly matching labels highlighted green, incorrectly mis-matching labels highlighted pink. NEV = Melanocytic nevus. BCC = Basal cell carcinoma. SCC = Squamous cell carcinoma. ACK = Actinic keratosis. BEK = Benign keratosis. DFB = Dermatofibroma. MEL = Melanoma. VAS = Vascular lesion.  }
    \label{fig:derm}
\end{figure}

Important to note is the distinction between \ourmodel evaluations using a text-prompt classification head, which essentially acts as a ``black-box'' classifier using softmax, or using k-nearest neighbor image-image search as the classification (or regression) mechanism, which is inherently a transparent and interpretable approach that is able to yield evidence behind decisions. In many cases image-image search performs at a similar level of mAUC or better, demonstrating that building inherently interpretable methods can come without performance penalties in these situations. A visual example of image-image search on the ISIC2019 dataset is shown in Fig. ~\ref{fig:derm}.

For each of the tasks reported in Table ~\ref{table:comprehensive_results}, a category level breakdown of performance is available in the Appendix (Section ~\ref{sec:categoryperf}). 


The datasets leveraged for this study provide limited access to demographic information. Where gender information was provided, we stratified performance according to the specifications. The results are shown in Table ~\ref{tab:rai}. For chest X-ray classification in LTCXR, mAUC were nearly identical within 1 point. For RSNA BoneAge, absolute L1 error by months was within 0.3 months.

\subsection{Task Fine-Tuning Experiments}

\begin{table}[t!]
  \centering
  \begin{threeparttable}
    \caption{AUC Fine-tuning Results on 3 Categories of ChestX-ray14}
    \label{tab:cxr14}
    \begin{tabular}{@{}lccccc@{}}
      \toprule
      Model & \# Eval Domains & \# Params & Edema & Consolidation & Effusion \\
      \midrule 
      \ourmodel \tnote{a} & 14 & 360 M & \textbf{0.860} & 0.744 & 0.832 \\
      CXR Foundation~\cite{cxrfoundation}\tnote{b} &1 & 240 M & 0.827 & 0.727 & 0.813 \\
      ELIXR~\cite{elixr} \tnote{b,c} &1 & 240 M& 0.856 & \textbf{0.749} & \textbf{0.835} \\
      \bottomrule
    \end{tabular}
    \begin{tablenotes}
      \small
      \item[a] The \ourmodel model was evaluated on 13,980 images not part of LTCXR training.
      \item[b] The comparison models were evaluated on full 25,596 test set.
      \item[c] ELIXR numerical performance is estimated from a visual graph (supplemental Fig. 2 of reference) using a ruler. 
    \end{tablenotes}
  \end{threeparttable}
\end{table}

Fine-tuning results are shown in Table ~\ref{tab:cxr14}, and compared to results published from CXR Foundation~\cite{cxrfoundation} and ELIXR~\cite{elixr}. Across the 3 categories studied, \ourmodel is able to reproduce performance of ELIXR to within a half a point in AUC, while being trained on 14X as many domains, using only 1.5X more parameters (120M).  

\subsection{Report Generation}

\begin{table}[t!]
  \centering
  \setlength{\extrarowheight}{1.0pt}  
  \renewcommand{\arraystretch}{1.1} 
  \small
  \begin{threeparttable}
    \caption{Report Generation Performance on MIMIC-CXR}
    \label{tab:report_gen}
    \setlength{\tabcolsep}{0.7mm}{    
        \begin{tabularx}{\textwidth}{lcccccccccccccc}
            \toprule
            \multirow{4}{*}{Model}& \multicolumn{8}{c}{CheXbert} &  \multicolumn{2}{c}{\multirow{3}{*}{RadGraph}}  & \multicolumn{2}{c}{\multirow{3}{*}{BLEU}} & \multirow{3}{*}{ROUGE}\\ \cmidrule(){2-9}
            
            & \multicolumn{4}{c}{(``uncertain'' as \emph{negative})} & \multicolumn{4}{c}{(``uncertain'' as \emph{positive})} &  \\ \cmidrule(r){2-5} \cmidrule(l){6-9} %
            & \multicolumn{2}{c}{Micro-avg} & \multicolumn{2}{c}{Macro-avg} & \multicolumn{2}{c}{Micro-avg} & \multicolumn{2}{c}{Macro-avg}  \\ %
            & F1-14 & F1-5 & F1-14 & F1-5 & F1-14 & F1-5 & F1-14 & F1-5 & F1 & ER & (1) & \multicolumn{1}{c}{(4)} & \multicolumn{1}{c}{(L)}   \\ \midrule %

            \multicolumn{3}{l}{\cellcolor{gray!25}\textit{Single Image. Model size \textgreater 7B}}\\ 
            
            LLaVA-Rad \cite{llavarad} \tnote{F} & \textbf{57.3} & 57.4 & 39.5 & 47.7 & \textbf{57.3} & \textbf{60.2} & \textbf{44.0} & \textbf{53.3} & - & 29.4 & 38.1 & 15.4 & 30.6 \\ %
            Med-PaLM M \cite{medpalm}& 53.6 & \textbf{57.9} & \textbf{39.8} & \textbf{51.6} & - & - & - & - & \textbf{26.7} & - & 32.3 & 11.3 & 27.3 \\ %
            GPT-4V & 35.5 & 25.8 & 20.4 & 19.6 & 35.6 & 33.3 & 25.3 & 29.6 & - & 13.2 & 16.4 & 1.9 & 13.2 \\ %
            GPT-4o finetune\tnote{d} & 48.9 & 52.7 & 33.0 & 43.8 & 49.3 & 54.7 & 36.7 & 47.8 & 26.4 & 29.3 & \textbf{39.6} & \textbf{17.8} & 32.1 \\ %
            GPT-4o-mini finetune\tnote{d} & 47.6 & 51.8 & 30.8 & 42.0 & 48.1 & 54.1 & 34.6 & 46.4 & 26.0 & 29.1 & 37.0 & 16.2 & \textbf{32.2} \\ %
            MAIRA-1 \cite{maira1} \tnote{F}& 55.7 & 56.0 & 38.6 & 47.7 & 55.3 & 58.8 & 42.3 & 51.7 & 24.3 & \textbf{29.6} & 39.2 & 14.2 & 28.9 \\ %
            CheXagent \cite{chen2024chexagent}  & 39.3 & 41.2 & 24.7 & 34.5  & 39.4 & 42.1 & 27.3 & 35.8 & - & 20.5 & 16.9 & 4.7 & 21.5 \\ %
            LLaVA-Med \cite{llava-med} \tnote{F} & 27.2 & 22.0 & 15.5 & 16.6 & 27.3 & 24.4 & 18.7 & 20.5 & - & 6.5 & 22.2 & 1.0 & 13.3 \\ %
            LLaVA \cite{llava} \tnote{F}& 22.9  & 23.4 & 15.4 & 17.5 & 23.7 & 26.9 & 17.0 & 20.3 & - & 4.5 & 21.0 & 1.3 & 13.8 \\ %
            \midrule
            \multicolumn{3}{l}{\cellcolor{gray!25}\textit{Multi-Image. Model size \textgreater 7B}}\\ 
            MAIRA-2 \cite{maira2} \tnote{c} & \textbf{58.1} & \textbf{59.1} & \textbf{41.6} & \textbf{50.4} & \textbf{58.5} & \textbf{61.3} & \textbf{45.8} & \textbf{54.6} & \textbf{34.6} & \textbf{39.6} & \textbf{46.0} & \textbf{23.1} & \textbf{38.4} \\
            GPT-4o finetune\tnote{c,d} & {52.8} & {54.0} & {32.7} & {43.7} & {52.2} & {55.2} & {36.3} & {47.1} & 32.9 & {36.5} & {40.3} & {19.2} & {37.7} \\
            \midrule
            \midrule
            \multicolumn{3}{l}{\cellcolor{gray!25}\textit{Single Image. Model size \textless 1B}}\\ 
            CvT2Dist. \cite{nicolson_improving_2023} & \textbf{44.2} & - & \textbf{30.7} & - & - & - & - & - & - & - & \textbf{39.3} & \textbf{12.7} & \textbf{28.6} \\ %
            $\mathcal{M}^2$ trans \cite{m2trans} & - & - & - & - & - & \textbf{56.7} & - & - & - & - & - & 11.4 & - \\ %
            RGRG \cite{rgrg} \tnote{F} & - & - & - & - & - & 54.7 & - & - & - & - & 37.3 & 12.6 & 26.4 \\ %
            R2Gen \cite{chen-emnlp-2020-r2gen} & - & - & - & - & 22.8 & 34.6 & - & - & - & - & 35.3 & 8.6 & 27.7 \\ %
            TieNet \cite{wang2018tienet}  & - & - & - & - & - & 27.1 & - & - & - & - & - & 8.1 & - \\ %
            \midrule
            \midrule
            \multicolumn{3}{l}{\cellcolor{gray!25}\textit{Ours -- Single Image. Model size 0.4B}}\\ 
            \ourmodel \tnote{a}  & \textbf{53.6} & \textbf{57.2} & \textbf{37.0} & \textbf{48.3} & \textbf{53.3} & \textbf{58.6} & \textbf{40.7} & \textbf{51.7} & \textbf{25.5} & \textbf{28.8} & \textbf{37.0} & \textbf{15.2} & \textbf{31.6} \\ %
            Ablative Baseline \tnote{b} & 49.1 & 53.9 & 31.0 & 43.9 & 49.7 & 56.2 & 35.2 & 49.2 & 23.6 & 27.1 & 33.5 & 13.4 & 29.9 \\ %
            \midrule
            \ourmodel \tnote{a,F}  & \textbf{56.3} & \textbf{57.9} & \textbf{38.4} & \textbf{49.3} & \textbf{55.7} & \textbf{59.3} & \textbf{43.2} & \textbf{52.1} & \textbf{25.4} & \textbf{28.5} & \textbf{37.3} & \textbf{15.3} & \textbf{31.7} \\ %
            Ablative Baseline \tnote{b,F} & 52.4 & 55.2 & 32.2 & 45.6 & 52.9 & 58.0 & 36.5 & 50.8  & 26.9  & 23.6 & 33.6 & 13.4 & 29.9\\ %
            
            \bottomrule
        \end{tabularx}
    }
    \begin{tablenotes}
      \small
      \item[*] CheXbert \cite{smit2020chexbert} and RadGraph (F1 \& ER) metrics \cite{jain2021radgraph} are Clinical Efficacy (CE) metrics for report generation. We use results from \cite{llavarad} if not available in the original papers. BLEU and ROUGE may be implemented differently in existing work; we use the official code from COCO Captions \cite{coco_captions}      
      \item[a] Report generation model is built by connecting \ourmodel with a 6-layer transformer text decoder.
      \item[b] An ablation study which uses the same model architecture but not \ourmodel's weights
      \item[c] The MAIRA-2 MIMIC-CXR benchmark has been redesigned to better reflect clinical scenarios by combining multiple image instances from the same case into a single instance. Therefore, direct comparisons to other approaches evaluated on the single image benchmark cannot be made
      \item[d] GPT finetuning experiments leveraged the MIMIC-CXR dataset only. Other models used more data.
      \item[F] The testing set includes only frontal-view images.
      
    \end{tablenotes}
  \end{threeparttable}
\end{table}

In Table ~\ref{tab:report_gen}, we present the results on the official test split of MIMIC-CXR, currently the largest report generation dataset. We include both lexical metrics (e.g. BLEU) and clinical efficacy metrics (e.g. CheXbert). Thanks to the feature representation from \ourmodel, our model achieves comparable performance with the latest state-of-the-art methods, but with only 7\% of the model size. To quantify the gain from using \ourmodel weights, we conducted an ablation study replacing \ourmodel with a randomly initialized image encoder of the same architecture that is trained in a pretraining stage. \ourmodel outperforms this ablative baseline, which demonstrates that \ourmodel provides a useful initialization that benefits report generation performance.

Finetuning GPT-4o and GPT-4o-mini led to new SOTA performance in lexical metrics (BLEU and ROUGE) for the single image MIMIC findings generation, but underperformed \ourmodel in terms of clinical CheXbert metrics. 

\subsection{Independent Clinical Evaluation}
\label{res:siteevaluation}

\begin{figure}[h!]
    \centering
    \includegraphics[width=0.92\textwidth]{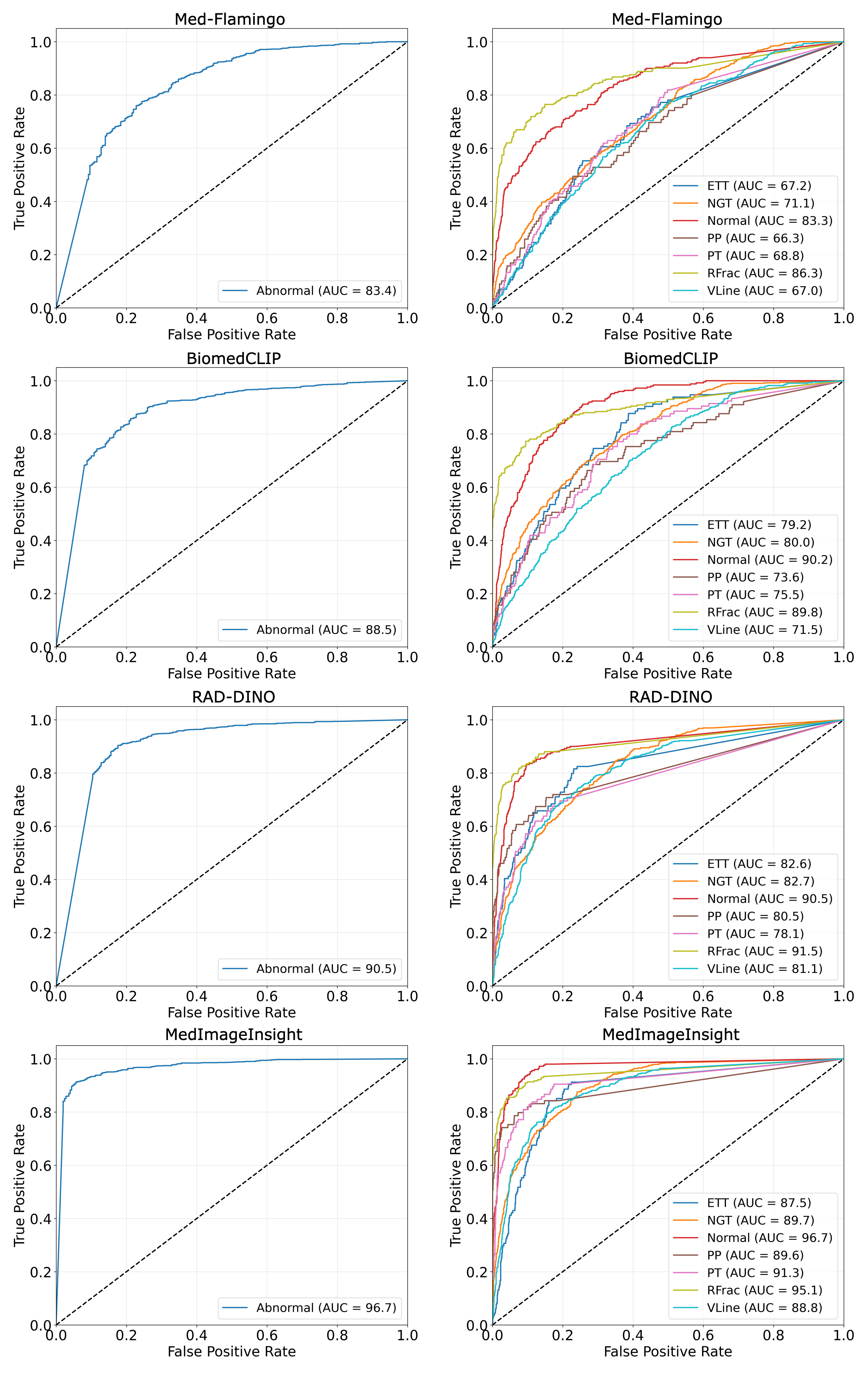}
    \caption{ROC curves from independent site evaluation using KNN classification on extracted image features for binary and multi-class image classification. \textit{Left column:} ROC curves for binary classification. \textit{Right column:} ROC curves for multi-class classification.}
    \label{fig:roc}
\end{figure}

AUC scores from an independent evaluation for binary and multi-class classification on our clinical partner dataset are presented in Table~\ref{tab:auc}, with ROC curves shown in Fig.~\ref{fig:roc}. \ourmodel achieved a higher AUC score than Med-Flamingo, BiomedCLIP, and RAD-DINO in all categories, demonstrating the utility of the image feature representations of \ourmodel for no-training image-image search based classification on real-world unseen data. Importantly, \ourmodel also exhibits robust performance for fairness across gender and age demographics, as detailed in the Appendix (Table~\ref{tab:fairness_gender} and Table~\ref{tab:fairness_age}). \ourmodel outperforms other models in both female and male subcategories, with the performance on females marginally higher than on males (by 0.3 points mAUC). Similarly, across various age demographic groups, \ourmodel demonstrates consistent high performance, with minimal variations in mAUC compared to other models. These results suggest \ourmodel provides equitable outcomes across age and gender groups in this task.

\begin{table}[h!]
  \centering
  \setlength{\extrarowheight}{1.0pt}  
  \small
  \begin{threeparttable}
    \caption{AUC Results from Independent Clinical Evaluation using KNN Classification}
    \label{tab:auc}
    \setlength{\tabcolsep}{2.0mm}{    
        \begin{tabularx}{\textwidth}{lccccccccc}
            \toprule
            \multirow{3}{*}{Model} & \multicolumn{8}{c}{Multi-class } & \multicolumn{1}{c}{Binary-class} \\ \cmidrule(r){2-9} \cmidrule(l){10-10} %
            & ETT &  NGT &  PP & PT & RFrac &  VLine &  Normal & mAUC & Abnormal  \\ \midrule 
            Med-Flamingo \cite{MedFlamingo23} & 67.2 & 71.1 & 66.3 & 68.8 & 86.3 & 67.0 & 83.3 & 72.9 & 83.4  \\
            BiomedCLIP \cite{PMC15M} & 79.2 & 80.0 & 73.6 & 75.5 & 89.8 & 71.5 & 90.2 & 80.0 & 88.5 \\
            RAD-DINO \cite{fernando2024raddino} & 82.6 & 82.7 & 80.5 & 78.1 & 91.5 & 81.1 & 90.5 & 83.9 & 90.5 \\
            \ourmodel & \bf 87.5 & \bf 89.7 & \bf 89.6 & \bf 91.3 & \bf 95.1 & \bf 88.8 & \bf 96.7 & \bf 91.2 & \bf 96.7 \\
            \bottomrule
        \end{tabularx}
    }
    \begin{tablenotes}
      \footnotesize
      \item[*] ETT=endotracheal tube, NGT=nasogastric tube, PP=Pneumoperitoneum, PT=pneumothorax, RFrac=rib fracture, VLine=vascular lines
    \end{tablenotes}
  \end{threeparttable}
\end{table}

\subsection{3D Medical Image Retrieval}

We evaluate 3D image retrieval methods on four datasets from the 3D-MIR benchmark: Liver, Pancreas, Lung, and Colon \cite{3dmir2023}. We assess 3D search results when using four different models for embeddings generation: BiomedCLIP \cite{PMC15M}, Med-Flamingo \cite{MedFlamingo23}, BiomedGPT \cite{BiomedGPT24}, and \ourmodel. 

We compute Precision@k (P@k) and Average Precision (Avg Prec.) by comparing the tumor flag and the tumor stage of the query volume and the top-k retrieved volumes. We present the retrieval performance results in Table~\ref{tab:3dmir}. 

\ourmodel-based retrieval outperforms solutions based on two other models, Med-Flamingo and BiomedCLIP, in tumor presence matching on the four datasets. In tumor stage matching, the \ourmodel-based search outperforms the three other models (Med-Flamingo, BiomedCLIP, and BiomedGPT) in Average Precision in two out of four datasets, and Precision@5 in three out of four datasets (Pancreas, Lung, and Colon). 

When considering the macro-average across all organs/datasets, \ourmodel-based retrieval outperforms substantially all other models in all evaluation metrics (Tumor Presence/Staging Precision@k and Average Precision).

\begin{table}[h!]
  \centering
  \setlength{\extrarowheight}{1.0pt}  
  \small
  \begin{threeparttable}
    \caption{3D Image Retrieval Performance on the 3D-MIR Benchmark \cite{3dmir2023}: Liver, Pancreas, Lung, and Colon Datasets}
    \label{tab:3dmir}
    \setlength{\tabcolsep}{2.7mm}{    
        \begin{tabularx}{\textwidth}{lccccccccc}
            \toprule
            \multirow{3}{*}{Model} & \multicolumn{4}{c}{Tumor Presence} & \multicolumn{4}{c}{Tumor Staging} \\ \cmidrule(r){2-5} \cmidrule(l){6-9} %
            & P@3 &  P@5 &  P@10 & Avg Prec. & P@3 &  P@5 &  P@10 & Avg Prec.  \\ \midrule 
            & \multicolumn{8}{c}{Liver} \\ \cmidrule(){2-9}
            Med-Flamingo \cite{MedFlamingo23} & 42.1 & 45.3 & 47.7 & 50.6 & 35.1 & 35.8 & 36.8 & 41.4  \\
            BiomedCLIP \cite{PMC15M} & 79.0 & 77.9 & 76.9 & 81.6 & 40.4 & 43.2 & 44.6 & 50.2 \\
            BiomedGPT \cite{BiomedGPT24} & \bf 82.5 & \bf 81.0 & 79.6 & \bf 84.0 & \bf 45.6 & \bf 49.5 & \bf 46.1 & \bf 55.4 \\
            \ourmodel & 80.7 & 77.9 & \bf 81.7 & 82.3 & 43.9 & 42.1 & 44.5 & 54.6 \\
            \midrule
             & \multicolumn{8}{c}{Pancreas} \\ \cmidrule(){2-9}
            Med-Flamingo \cite{MedFlamingo23} & 94.8 & 93.1 & 92.3 & 94.9 & \bf 64.6 & \bf 58.1 & 53.6 & 65.6 \\
            BiomedCLIP \cite{PMC15M} &  97.9 & 97.5 & 96.9 & 97.9 & 47.9 & 50.6 & 56.6 & 61.9 \\ 
            BiomedGPT \cite{BiomedGPT24} &  \bf 100 & \bf 100  & \bf 99.4 & \bf 100 & 44.8 & 51.9 & \bf 57.8 & 61.1 \\
            \ourmodel & \bf 100 & \bf 100 & \bf  99.4 & 99.9 & 60.4 & \bf 58.1 & 57.6 & \bf 67.3 \\
            \midrule
            & \multicolumn{8}{c}{Lung} \\ \cmidrule(){2-9}
            Med-Flamingo \cite{MedFlamingo23} & 96.9 & 96.9 & 95.5 & 97.3 & 70.8 & 69.4 & 66.8 & 75.6 \\
            BiomedCLIP \cite{PMC15M} & 99.3  & 98.3 & \bf 98.3 & 99.1 & \bf 73.7 & 65.1 & 61.8 & \bf 83.1 \\
            BiomedGPT \cite{BiomedGPT24} &  \bf 100 & \bf 100 & 94.1 &  \bf 100 & 72.0 & 72.9 & 66.2 & 76.2   \\
            \ourmodel & \bf 100 & \bf 100 & 96.9 &  99.7 & 69.8 & \bf 73.8 & \bf 68.6 & 76.1 \\
            \midrule
            & \multicolumn{8}{c}{Colon} \\ \cmidrule(){2-9}
            Med-Flamingo \cite{MedFlamingo23} & 54.2 & 60.00 & 53.9 & 69.6 & 36.1 & 42.5 & 36.1 & 54.9 \\
            BiomedCLIP \cite{PMC15M} & 84.7 & 83.3 & 78.8 & 87.1 & 51.4 & 51.7 & 47.9 & 63.3 \\
            BiomedGPT \cite{BiomedGPT24} &  59.7 & 52.5 & 48.3 & 80.5 & 50.0 &  46.7 & 44.4 & 59.6  \\
            \ourmodel & \bf 100 & \bf  99.2 & \bf 97.3 & \bf 99.7 & \bf 65.3 & \bf 62.5 & \bf 60.6 & \bf 69.4\\

            \midrule
            & \multicolumn{8}{c}{Average across all organs} \\ \cmidrule(){2-9}
            Med-Flamingo \cite{MedFlamingo23} &  72.0 & 73.8 & 72.3 & 78.1 & 51.7 & 51.4 & 48.3 & 59.3 \\
            BiomedCLIP \cite{PMC15M} & 90.2 & 89.3 & 87.7 & 91.4 & 53.4 & 52.6 & 52.7 & 64.6  \\
            BiomedGPT \cite{BiomedGPT24} &  85.5 & 83.4 & 80.4 & 91.1 & 53.1 & 55.2 & 53.6 & 63.1 \\
            \ourmodel &  \bf 95.2 & \bf 94.3 & \bf 93.8 & \bf  95.4 & \bf 59.8 & \bf 59.1 & \bf 57.8 & \bf 66.9  \\
            \bottomrule
        \end{tabularx}
    }
  \end{threeparttable}
\end{table}

\section{Methods}
\label{methods}

\subsection{Pre-training}

\begin{figure}[t!]
    \centering
    \includegraphics[width=0.9\textwidth]{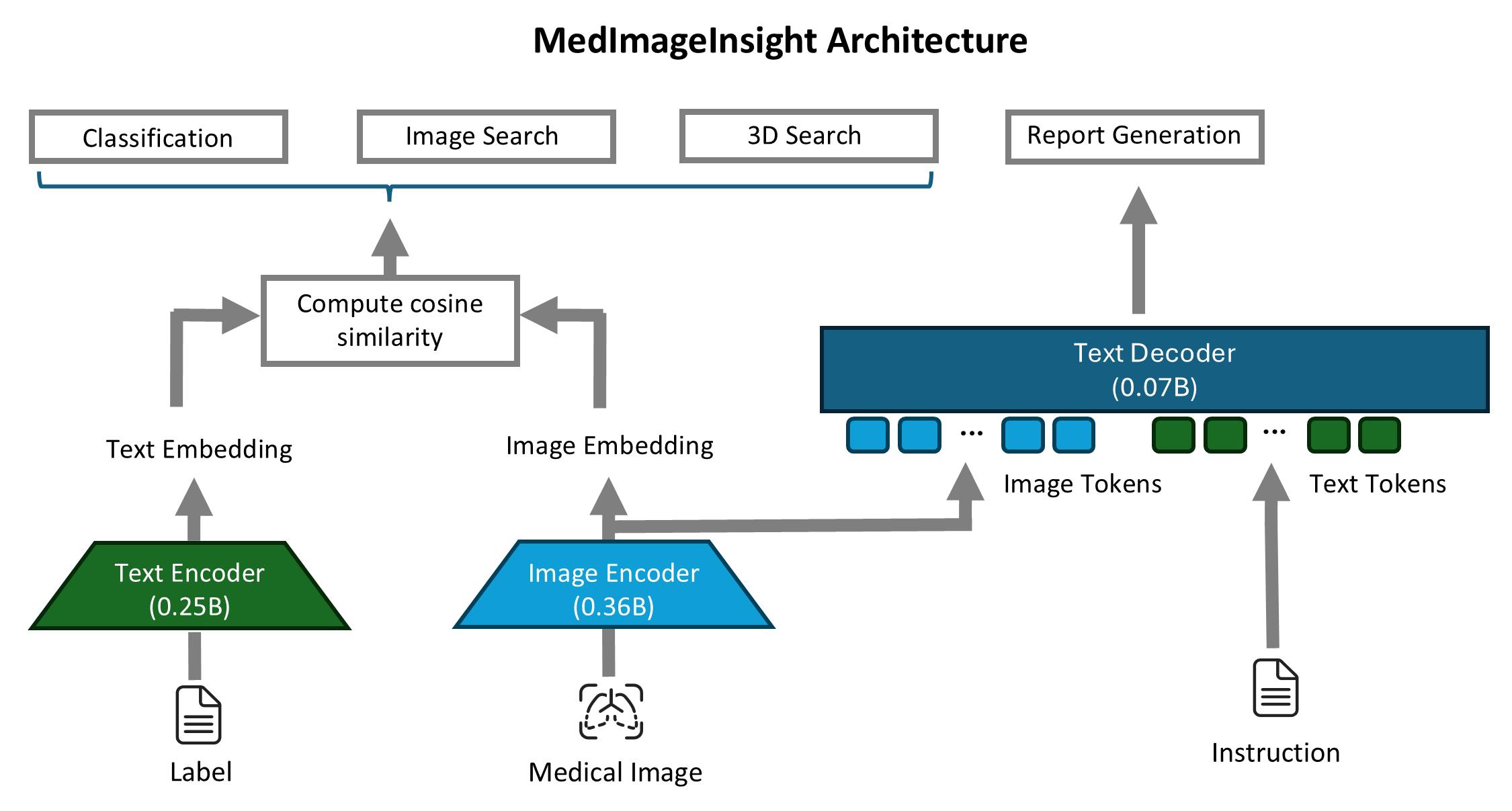}
    \caption{Overview of the \ourmodel foundation model architecture. UniCL is used as the pre-training objective function. }
    \label{fig:arch}
\end{figure}

\ourmodel was derived from the Florence computer vision foundation model~\cite{florence}. The overall architecture of \ourmodel is shown in Fig.~\ref{fig:arch}. \ourmodel is a two-tower architecture similar to CLIP~\cite{clip}, however in this implementation the image encoder employs the DaViT architecture~\cite{davit}, and UniCL~\cite{unicl} is used as the objective function. The DaViT image encoder used in \ourmodel is 360M parameters and the language encoder is 252M parameters. Importantly, for classification, the language encoder is run once in inference to generate a classifier head to classify N individual image instances. Optionally, the vision encoder can be connected to a decoder for report generation.

\textbf{Training Setup:} We trained our model using the UniCL objective with with mixed precision (FP16) enabled. The training employed the AdamW optimizer with a learning rate of 1E-5 and weight decay of 0.2. The batch size was set to 1024, distributed across multiple V100 GPUs, with gradient accumulation steps computed based on the batch size and number of GPUs. The learning rate followed a cosine decay schedule with warmup steps. Additionally, gradient clipping at 1.0 was used to ensure stability. For data augmentation we used random rescale and ratio variations, as well as random erasing, from {\em torchvision.transforms}.  Training was run for 40k iterations.

\textbf{Carbon Footprint:} Pretraining utilized a cumulative 7680 GPU hours of computation on hardware of type V100 (TDP of 250W-400W). Estimated total emissions were 0.89184 tCO2eq. We trained on Azure Machine Learning. We used 64 V100 GPUs. Compute region was West US 2.

\subsubsection{Medical Pre-training Datasets}

Datasets used for model pre-training span over 14 medical imaging domains and are summarized in Fig. ~\ref{fig:overview} and Table ~\ref{tab:pretraining}. Datasets include MIMIC-CXR~\cite{mimic}, NIH-CXR-LT~\cite{ltcxr}, IRMA 2009~\cite{irma2009}, RSNA BoneAge~\cite{rsnaboneage}, GP BoneAge Atlas~\cite{gpboneage}, UPENN~\cite{upenn}, TCGA~\cite{tcgasarc}, SD198~\cite{sd198}, ISIC2019~\cite{isic2017,isic2018,isic2019}, PatchCamelyon~\cite{pcam}, RSNA Mammography~\cite{rsnamammography}, LIDC-IDRI~\cite{lidcidri}, PAD-UFES-20~\cite{padufes20},  ODIR-5K~\cite{odir5k}, OCT2017~\cite{oct2017}, Breast Ultrasound Image (BUSI)~\cite{busi}, HMC-QU~\cite{hmcqu}, Medical Segmentation Decathlon (MSD)~\cite{msd}, ROCO~\cite{roco}, PMC-15M~\cite{PMC15M}, and partner and proprietary datasets. For full details about the data leveraged for pre-training, evaluation, and pre-processing steps, please refer to Section ~\ref{sec:appendixdatasets}.

\begin{table}[t!]
  \centering
  \begin{threeparttable}
    \caption{\ourmodel Pre-training \& Classification/Search Evaluation Datasets}
    \label{tab:pretraining}
    \begin{tabular}{@{}lllll@{}}
      \toprule
      Dataset & Modalities* & Format & \begin{tabular}{@{}c@{}}Data Used for\\Pretraining\end{tabular} & \begin{tabular}{@{}c@{}}Data Used for\\Evaluation\end{tabular} \\
      \midrule 
        MIMIC-CXR ~\cite{mimic} & CXR & Image-Text & 203,170 & 0 \\
        NIH-CXR-LT ~\cite{ltcxr} & CXR & Image-Label (20) & 68,058 & 20,279  \\
        IRMA 2009 ~\cite{irma2009} & XR & Image-Label (137) & 12,677 & 1,733 \\
        RSNA BoneAge ~\cite{rsnaboneage} & XR & Image-Age & 12,611 & 1,425 \\
        GP BoneAge Atlas ~\cite{gpboneage} & XR & Image-Age & 59 & 0 \\
        UPENN ~\cite{upenn} & MRI & Image-Text & 4,645 & 0 \\
        TCGA ~\cite{tcgasarc} & MRI/CT & Image-Text & 5,643 & 0 \\
        SD198 ~\cite{sd198} & CP & Image-Label (198) & 3,253 & 3,331 \\
        ISIC2019 ~\cite{isic2017,isic2018,isic2019} & DM & Image-Label (8) & 20,268 & 5,063 \\
        PatchCamelyon ~\cite{pcam} & Hist. & Image-Label (2) & 262,144 & 32,768  \\
        RSNA Mammography ~\cite{rsnamammography} & XR & Image-Label (4) & 43,764 & 10,942 \\
        LIDC-IDRI ~\cite{lidcidri} & CT & Image-Label (4) & 80,201 & 4,574 \\
        PAD-UFES-20 ~\cite{padufes20} & CP & Image-Label (6) & 2,065 & 233 \\
        ODIR-5K ~\cite{odir5k} & Fundus & Image-Label (79/28)\textsuperscript{a} & 5,228 & 1,267 \\
        OCT2017 ~\cite{oct2017} & OCT & Image-Label (4) & 83,484 & 968 \\
        OCT2018 ~\cite{oct2017b} & OCT & Image-Label (4) & 0 & 1000 \\
        BUSI ~\cite{busi} & US & Image-Label (3) & 623 & 157 \\
        HMC-QU ~\cite{hmcqu} & US & Image-Label (2) & 4,714 & 2,132 \\
        Bing Echo View & US & Image-Label (7) & 100 & 54 \\
        MSD ~\cite{msd}& CT & Image-Text & 116,500 & 0 \\
        Bing Exam Parameter & CT/MRI & Image-Label (21) & 1202 & 401 \\
        ROCO ~\cite{roco} & Mixed & Image-Text & 70,308 & 0 \\
        MGB CXR & CXR & Image-Label (95/80)\textsuperscript{a} & 731 & 684 \\
        MGB BoneAge & XR & Image-Age & 0 & 120 \\
        PMC-15M ~\cite{PMC15M} & Mixed & Image-Text & 952,556 & 0 \\
        Proprietary & Mixed & Image-Text/Label & 1,832,206 & 0 \\
        \midrule 
        Total & Mixed & Image-Text/Label/Age & 3,786,210 & 86,131 \\

      \bottomrule
    \end{tabular}
    \begin{tablenotes}
      \small
      \item[*] CXR = Chest X-Ray, XR = Other X-Ray, MRI = Magnetic Resonance Imaging, CT = Computed Tomography, CP = Clinical Photography (Dermatology), Hist. = Histology, DM = DerMoscopy (Dermatology), Fundus = Fundus Imaging, US = Ultrasound, OCT = Optical Coherence Tomography 
      \item[a] Train/test labels. More labels were available for training than testing due to dataset distributions.  
    \end{tablenotes}
  \end{threeparttable}
\end{table}

\subsection{Evaluation}

We evaluate \ourmodel on a wide range of medical imaging domains, across a variety of tasks, including base image-text and image-image search (Section ~\ref{sec:basemethods}), task fine-tuning (Section ~\ref{sec:finetunemethods}), report generation (Section ~\ref{sec:reportgenmethods}), an independent site evaluation of KNN classification for chest X-rays (Section ~\ref{sec:siteevaluation}), and 3D medical image retrieval (Section ~\ref{sec:3dmirmethods}). 

Image-text search in the context of a two-tower architecture is analogous to classification. Image-image search is analogous to KNN classification, an inherently interpretable evidence-based classification approach. Task fine-tuning demonstrates the ability of the model to further specialize to specific tasks. 3D medical image retrieval is essentially a 3D form of image-image search.

\subsubsection{Base Image-Text and Image-Image Search}
\label{sec:basemethods}

As a two-tower architecture optimized with a UniCL objective function, \ourmodel supports 3 primary modes of inference operation: 1) Text-Image Search, 2) Image-Text Search, and 3) Image-Image Search. For the purposes of this work, we focus on Image-Text Search (which is a classification task with a classification head defined by text embeddings) and Image-Image Search (which is KNN classification). Importantly, Image-Image Search is inherently a transparent and ``explainable'' mode of operation, as the system is able to provide direct evidence for classification decisions. In Image-Image search mode, the instances from the original training datasets serve as search databases. For all Image-Image search classification tasks, we use k = 20 with weighted voting (the weight defined as the dot product between query and result). For all regression tasks (bone age datasets), we use k = 100, weighted voting, and predicting the age with the highest score.


On datasets that provide stratification according to patient demographics, we assessed performance breakdown according to these categories. Two datasets supported this analysis by gender: LTCXR, and RSNA BoneAge.  

\subsubsection{Task Fine-Tuning Experiments}
\label{sec:finetunemethods}

To evaluate the adaptability and performance of \ourmodel in specific medical imaging tasks, we conducted fine-tuning experiments on the ChestX-ray14 dataset. This dataset is widely used in the medical AI community and contains 112,120 frontal-view chest X-ray images from 30,805 unique patients, with 14 disease labels. We focused on three clinically important conditions from the ChestX-ray14 dataset: Edema, Consolidation, and Effusion. These 3 are selected because they enable us to compare \ourmodel with two other state-of-the-art models: CXR Foundation ~\cite{cxrfoundation} and ELIXR ~\cite{elixr}.

For fine-tuning we froze early layers in both encoders. Specifically, for the vision encoder we froze the first stage. For the text encoder, we froze positional and token embeddings as well as the first 3 blocks of the text encoder. Learning rate was 1E-6, batch size 1024, weight decay 0.4, and training was conducted for 800 iterations.

\subsubsection{Report Generation}
\label{sec:reportgenmethods}

We evaluate \ourmodel as the image backbone for radiology report generation. To do so, we connect it with a text decoder, which is a randomly initialized six-layer transformer with a feature dimension of 768. Following existing work \cite{llava-med,llavarad,maira1}, the inputs to the text decoder include the image patches and an instruction which describes the task. Image patches are projected to input tokens with a single linear layer. The instruction includes the ``Indication'' or ``History'' section of the report to provide the goal of the radiology study. Image and text tokens are concatenated and go through self-attention. 

The model was trained in two stages. In stage 1, we pre-train the model to align the image encoder and the text decoder. The \ourmodel image encoder is frozen and only the text decoder is updated (however, in the ablation study, the image encoder was randomly initialized and was trainable in this step). We use the CXR-697K dataset to be consistent with LLaVA-Rad \cite{llavarad}. The pre-training takes 400 epochs with a learning rate of 2E-5 and a batch size of 2048. In stage 2, the full model was fine-tuned on the MIMIC-CXR dataset. The fine-tuning takes 40 epochs, with a learning rate of 1E-5 for the backbone and 5E-5 for the text decoder, and a batch size to 512. The image size is 480x480. We set the maximum input and output text length to 710 and 256, respectively, and instruction is only used during fine-tuning.

As a baseline for our experiments, we include GPT-4o and GPT-4o-mini finetuned on the MIMIC-CXR data alone, using a private finetuning API in Azure. Training parameters were 10 epochs, with a learning rate multiplier of 2, and a batch size of 600.

\subsubsection{Independent Clinical Evaluation}
\label{sec:siteevaluation}

\ourmodel was evaluated on a private X-ray dataset from the University of Wisconsin-Madison, consisting of 8,842 images. The dataset was considered for both binary classification (normal study vs abnormal study) and multi-class classification, with categories including pneumothorax (PT), endotracheal tube (ETT), nasogastric tube (NGT), vascular lines (VLine), rib fracture (RFrac), pneumoperitoneum (PP), and normal study. Datasets were manually labelled for each class by trained radiologists, and the labels were mutually exclusive. The dataset was split into a search database, validation, and test sets in a 64\%, 16\%, and 20\% ratio, respectively. KNN image search was facilitated using image feature representations extracted from \ourmodel on the database set and evaluated on the validation set for hyper-parameter selection (the number of neighbors used for voting). Finally, the AUC score was computed using the probabilities predicted by the KNN model, post-Softmax, on the test dataset. For comparison, Med-Flamingo, BiomedCLIP, and RAD-DINO were evaluated in the same manner. 

We also assessed the fairness of our model with respect to gender and age on the test dataset that included 709 females and 1059 males. For the gender fairness assessment, we evaluated performance metrics across these two groups. For age-related fairness, we organized patients into the following age groups for detailed analysis: from 0 to 20 years (164 patients), over 20 to 40 years (237 patients), over 40 to 60 years (570 patients), over 60 to 80 years (610 patients), and over 80 to 100 years (99 patients). Patients whose records indicated an age older than 100 years were excluded from the age fairness analysis. This stratification was designed to provide insights into the model's performance across various life stages.

\subsubsection{3D Medical Image Retrieval}
\label{sec:3dmirmethods}

We leverage the embeddings generated by \ourmodel (and other baseline models) to generate a representative embedding vector for a given 3D volume. We selected the median-based aggregation of the 2D slices embeddings after a preliminary investigation on a random subset of the training set using median, standard deviation, average, and max pooling. 

In a first offline step, we generated the 3D embeddings of each volume in the training set following the median aggregation method and indexed them in a vector database. The 3D search is then performed by generating the representative 3D embedding vector for the query image and retrieving the nearest vectors from the database according to the dot product. 

For evaluation, we use the volumes in the test set as test queries and retrieve the top-k most similar volumes using our search method. We then compare the tumor flag and the tumor stage of the query volume and the top-k retrieved volumes to compute Precision@k and Average Precision. 

\section{Discussion}
\label{discussion}

In this work, we have introduced \ourmodel, a foundation model for medical imaging. With under 0.6 B parameters, \ourmodel is lightweight and leverages modern encoder architectures that surpass the performance of baseline ViT in prior studies~\cite{davit}. Spanning a wide range of modalities, including X-ray, CT, MRI, mammography, dermatology, ultrasound, and pathology, \ourmodel achieves state-of-the-art (SOTA) or human-expert level performance on several datasets where comparisons are available. High mAUC (greater than 0.9) is attained in most of the remaining datasets. Additionally, critical features for regulatory compliance, such as ROC curve generation, and practical implementation, including evidence-based classification through image-image search, are supported. 

\ourmodel demonstrates the ability to operate with its pre-trained capabilities for classification and image-image search across multiple domains. In an independent clinical evaluation of image-image search for evidence-based classification in chest X-ray, \ourmodel outperformed every other publicly accessible foundation model studied. Fine-tuning allows \ourmodel to achieve SOTA-equivalent performance compared to other specialized models in single domains. Furthermore, integrating the model with decoder models provides robust report generation capabilities. Compared to fine-tuning GPT-4o and GPT-4o-mini, \ourmodel outperforms in report findings generation according to clinical metrics, but underperforms on lexical metrics where the GPT-4o models set a new SOTA for findings generation from single images. 

By releasing the model weights as open-source, we aim to empower the broader research community to accelerate advancements in the application of AI within medical imaging, including the enhancement of transparency and reproducibility, and the facilitation of access to cutting-edge technology, ultimately benefiting the global healthcare community. 

See our responsible AI statement \footnote{Microsoft believes Responsible AI is a shared responsibility. We have identified six principles and practices help organizations address risks, innovate, and create value: fairness, reliability and safety, privacy and security, inclusiveness, transparency, and accountability. When downloaded or used in accordance with our terms of service, developers should work with their supporting model team to ensure this model meets requirements for the relevant use case and addresses unforeseen product misuse. 

While testing the model with images and/or text, ensure the the data is PHI free and that there are no patient information or information that can be tracked to a patient identity. 

The model is not designed for the following use cases: 

\begin{itemize}
    \item Use as a diagnostic tool or as a medical device - Using information extracted by our service in diagnosis, cure, mitigation, treatment, or prevention of disease or other conditions, as a substitute of professional medical advice, diagnosis, treatment, or clinical judgment of a healthcare professional.
    \item Scenarios without consent for data - Any scenario that uses health data for a purpose for which consent was not obtained.
    \item Use outside of health scenarios - Any scenario that uses non-medical related image and/or serving purposes outside of the healthcare domain
\end{itemize}}.

\section{Acknowledgements}

We would like to thank our partners at Mass-General Brigham Hospital, Keith J. Dreyer and Mannudeep Kalra, for providing access to real-world data and sharing their clinical expertise. We would like to thank colleagues Lin Liang, Zhengyuan Yang, and Lijuan Wang for feedback and advice on report generation experiments. We would like to thank Paul Vozila, Thomas Soemo, Dinei Florencio, and Cha Zhang for advice and support. We would also like to thank Microsoft Azure AI for their assistance in data download, reformatting, and rule-based filtering for a subset of the data used in this work.

\appendix

\bibliography{iclr2023_conference}

\begin{thebibliography}{10}

\bibitem{growthstats}
GlobeNewswire, ``With 5.8\% cagr, medical imaging market size worth usd 56.53 billion in 2028.''

\bibitem{humanai}
P.~Tschandl, C.~Rinner, Z.~Apalla, G.~Argenziano, N.~Codella, A.~Halpern, M.~Janda, A.~Lallas, C.~Longo, J.~Malvehy, J.~Paoli, S.~Puig, C.~Rosendahl, H.~P. Soyer, I.~Zalaudek, and H.~Kittler, ``Human–computer collaboration for skin cancer recognition,'' {\em Nature Medicine}, vol.~26, pp.~1229–--1234, 2020.

\bibitem{gpt3}
T.~Brown, B.~Mann, N.~Ryder, M.~Subbiah, J.~D. Kaplan, P.~Dhariwal, A.~Neelakantan, P.~Shyam, G.~Sastry, A.~Askell, {\em et~al.}, ``Language {M}odels are {F}ew-{S}hot {L}earners,'' {\em Advances in neural information processing systems}, vol.~33, pp.~1877--1901, 2020.

\bibitem{gpt4}
J.~Achiam, S.~Adler, S.~Agarwal, L.~Ahmad, I.~Akkaya, F.~L. Aleman, D.~Almeida, J.~Altenschmidt, S.~Altman, S.~Anadkat, {\em et~al.}, ``G{PT}-4 {T}echnical {R}eport,'' {\em arXiv preprint arXiv:2303.08774}, 2023.

\bibitem{gpt4o}
OpenAI, ``G{PT}-4o,'' 2024.

\bibitem{gemini}
R.~Anil, S.~Borgeaud, Y.~Wu, J.-B. Alayrac, J.~Yu, R.~Soricut, J.~Schalkwyk, A.~M. Dai, A.~Hauth, K.~Millican, {\em et~al.}, ``Gemini: A {F}amily of {H}ighly {C}apable {M}ultimodal {M}odels,'' {\em arXiv preprint arXiv:2312.11805}, vol.~1, 2023.

\bibitem{msrgpt4}
Q.~Liu, S.~Hyland, S.~Bannur, K.~Bouzid, D.~Castro, M.~Wetscherek, R.~Tinn, H.~Sharma, F.~P{\'e}rez-Garc{\'\i}a, A.~Schwaighofer, P.~Rajpurkar, S.~Khanna, H.~Poon, N.~Usuyama, A.~Thieme, A.~Nori, M.~Lungren, O.~Oktay, and J.~Alvarez-Valle, ``Exploring the boundaries of {GPT}-4 in radiology,'' in {\em Proceedings of the 2023 Conference on Empirical Methods in Natural Language Processing} (H.~Bouamor, J.~Pino, and K.~Bali, eds.), (Singapore), pp.~14414--14445, Association for Computational Linguistics, Dec. 2023.

\bibitem{usmle}
D.~Brin, V.~Sorin, A.~Vaid, A.~Soroush, B.~S. Glicksberg, A.~W. Charney, G.~Nadkarni, and E.~Klang, ``Comparing {C}hat{GPT} and {GPT}-4 performance in {USMLE} soft skill assessments,'' {\em Scientific Reports}, vol.~13, no.~1, p.~16492, 2023.

\bibitem{medprompt}
H.~Nori, Y.~T. Lee, S.~Zhang, D.~Carignan, R.~Edgar, N.~Fusi, N.~King, J.~Larson, Y.~Li, W.~Liu, {\em et~al.}, ``Can {G}eneralist {F}oundation {M}odels {O}utcompete {S}pecial-purpose {T}uning? {C}ase {S}tudy in {M}edicine,'' {\em arXiv preprint arXiv:2311.16452}, 2023.

\bibitem{PMC15M}
S.~Zhang, Y.~Xu, N.~Usuyama, H.~Xu, J.~Bagga, R.~Tinn, S.~Preston, R.~Rao, M.~Wei, N.~Valluri, {\em et~al.}, ``Biomed{CLIP}: a multimodal biomedical foundation model pretrained from fifteen million scientific image-text pairs,'' {\em arXiv preprint arXiv:2303.00915}, 2023.

\bibitem{llava-med}
C.~Li, C.~Wong, S.~Zhang, N.~Usuyama, H.~Liu, J.~Yang, T.~Naumann, H.~Poon, and J.~Gao, ``L{L}ava-{M}ed: Training a {L}arge {L}anguage-and-{V}ision {A}ssistant for {B}iomedicine in {O}ne {D}ay,'' {\em Advances in Neural Information Processing Systems}, vol.~36, 2024.

\bibitem{llavarad}
J.~M.~Z. Chaves, S.-C. Huang, Y.~Xu, H.~Xu, N.~Usuyama, S.~Zhang, F.~Wang, Y.~Xie, M.~Khademi, Z.~Yang, H.~H. Awadalla, J.~Gong, H.~Hu, J.~Yang, C.~Li, J.~Gao, Y.~Gu, C.~Wong, M.-H. Wei, T.~Naumann, M.~Chen, M.~P. Lungren, S.~Yeung-Levy, C.~P. Langlotz, S.~Wang, and H.~Poon, ``Towards a clinically accessible radiology foundation model: open-access and lightweight, with automated evaluation,'' 2024.

\bibitem{maira1}
S.~L. Hyland, S.~Bannur, K.~Bouzid, D.~C. Castro, M.~Ranjit, A.~Schwaighofer, F.~P{\'e}rez-Garc{\'\i}a, V.~Salvatelli, S.~Srivastav, A.~Thieme, {\em et~al.}, ``M{AIRA}-1: A specialised large multimodal model for radiology report generation,'' {\em arXiv preprint arXiv:2311.13668}, 2023.

\bibitem{maira2}
S.~Bannur, K.~Bouzid, D.~C. Castro, A.~Schwaighofer, S.~Bond-Taylor, M.~Ilse, F.~P{\'e}rez-Garc{\'\i}a, V.~Salvatelli, H.~Sharma, F.~Meissen, {\em et~al.}, ``M{AIRA}-2: {G}rounded {R}adiology {R}eport {G}eneration,'' {\em arXiv preprint arXiv:2406.04449}, 2024.

\bibitem{cxrfoundation}
A.~B. Sellergren, C.~Chen, Z.~Nabulsi, Y.~Li, A.~Maschinot, A.~Sarna, J.~Huang, C.~Lau, S.~R. Kalidindi, M.~Etemadi, {\em et~al.}, ``Simplified transfer learning for chest radiography models using less data,'' {\em Radiology}, vol.~305, no.~2, pp.~454--465, 2022.

\bibitem{elixr}
S.~Xu, L.~Yang, C.~Kelly, M.~Sieniek, T.~Kohlberger, M.~Ma, W.-H. Weng, A.~Kiraly, S.~Kazemzadeh, Z.~Melamed, {\em et~al.}, ``E{LIXR}: {T}owards a general purpose {X}-ray artificial intelligence system through alignment of large language models and radiology vision encoders,'' {\em arXiv preprint arXiv:2308.01317}, 2023.

\bibitem{medgemini}
K.~Saab, T.~Tu, W.-H. Weng, R.~Tanno, D.~Stutz, E.~Wulczyn, F.~Zhang, T.~Strother, C.~Park, E.~Vedadi, {\em et~al.}, ``Capabilities of {G}emini {M}odels in {M}edicine,'' {\em arXiv preprint arXiv:2404.18416}, 2024.

\bibitem{medpalm}
T.~Tu, S.~Azizi, D.~Driess, M.~Schaekermann, M.~Amin, P.-C. Chang, A.~Carroll, C.~Lau, R.~Tanno, I.~Ktena, {\em et~al.}, ``Towards {G}eneralist {B}iomedical {AI},'' {\em NEJM AI}, vol.~1, no.~3, p.~AIoa2300138, 2024.

\bibitem{gloria}
S.-C. Huang, L.~Shen, M.~P. Lungren, and S.~Yeung, ``Gloria: A multimodal global-local representation learning framework for label-efficient medical image recognition,'' in {\em Proceedings of the IEEE/CVF International Conference on Computer Vision (ICCV)}, pp.~3942--3951, October 2021.

\bibitem{biomedgpt}
K.~Zhang, R.~Zhou, E.~Adhikarla, Z.~Yan, Y.~Liu, J.~Yu, Z.~Liu, X.~Chen, B.~D. Davison, H.~Ren, J.~Huang, C.~Chen, Y.~Zhou, S.~Fu, W.~Liu, T.~Liu, X.~Li, Y.~Chen, L.~He, J.~Zou, Q.~Li, H.~Liu, and L.~Sun, ``A generalist vision–language foundation model for diverse biomedical tasks,'' {\em Nature Medicine}, Aug. 2024.

\bibitem{whitehouse2023ai}
{The White House}, ``Fact sheet: {Biden-Harris} administration secures voluntary commitments from leading artificial intelligence companies to manage the risks posed by {AI},'' 2023.

\bibitem{pranavhumani}
N.~Agarwal, A.~Moehring, P.~Rajpurkar, and T.~Salz, ``Combining human expertise with artificial intelligence: Experimental evidence from radiology,'' 2024.

\bibitem{ltcxr}
G.~Holste, S.~Wang, Z.~Jiang, T.~C. Shen, G.~Shih, R.~M. Summers, Y.~Peng, and Z.~Wang, {\em Long-{T}ailed {C}lassification of {T}horax {D}iseases on {C}hest {X}-{R}ay: {A} {N}ew {B}enchmark {S}tudy}, p.~22–32.
\newblock Springer Nature Switzerland, 2022.

\bibitem{rsnaboneageexpert}
D.~B. Larson, M.~C. Chen, M.~P. Lungren, S.~S. Halabi, N.~V. Stence, and C.~P. Langlotz, ``Performance of a deep-learning neural network model in assessing skeletal maturity on pediatric hand radiographs,'' {\em Radiology}, vol.~287, 2018.

\bibitem{isic2019sota}
N.~Gessert, M.~Nielsen, M.~Shaikh, R.~Werner, and A.~Schlaefer, ``Skin lesion classification using loss balancing and ensembles of multi-resolution efficientnets,'' {\em ISIC 2019}, 2019.

\bibitem{sd198sota}
J.~Yang, X.~Sun, J.~Liang, and P.~L. Rosin, ``Clinical skin lesion diagnosis using representations inspired by dermatologist criteria,'' in {\em 2018 IEEE/CVF Conference on Computer Vision and Pattern Recognition}, pp.~1258--1266, 2018.

\bibitem{pcamsota}
S.~Graham, D.~Epstein, and N.~Rajpoot, ``Dense {S}teerable {F}ilter {CNN}s for {E}xploiting {R}otational {S}ymmetry in {H}istology {I}mages,'' 2020.

\bibitem{oct2017sota}
S.~A. Kamran, S.~Saha, A.~S. Sabbir, and A.~Tavakkoli, ``Optic-net: A {N}ovel {C}onvolutional {N}eural {N}etwork for {D}iagnosis of {R}etinal {D}iseases from {O}ptical {T}omography {I}mages,'' in {\em 2019 18th IEEE International Conference On Machine Learning And Applications (ICMLA)}, IEEE, Dec. 2019.

\bibitem{oct2017human}
D.~S. Kermany, M.~Goldbaum, W.~Cai, C.~C. Valentim, H.~Liang, S.~L. Baxter, A.~McKeown, G.~Yang, X.~Wu, F.~Yan, {\em et~al.}, ``Identifying medical diagnoses and treatable diseases by image-based deep learning,'' {\em cell}, vol.~172, no.~5, pp.~1122--1131, 2018.

\bibitem{chen2024chexagent}
Z.~Chen, M.~Varma, J.-B. Delbrouck, M.~Paschali, L.~Blankemeier, D.~Van~Veen, J.~M.~J. Valanarasu, A.~Youssef, J.~P. Cohen, E.~P. Reis, {\em et~al.}, ``Chexagent: Towards a foundation model for chest x-ray interpretation,'' {\em arXiv preprint arXiv:2401.12208}, 2024.

\bibitem{llava}
H.~Liu, C.~Li, Q.~Wu, and Y.~J. Lee, ``Visual instruction tuning,'' {\em Advances in neural information processing systems}, vol.~36, 2024.

\bibitem{nicolson_improving_2023}
A.~Nicolson, J.~Dowling, and B.~Koopman, ``Improving chest {X}-ray report generation by leveraging warm starting,'' {\em Artificial Intelligence in Medicine}, vol.~144, p.~102633, 2023.

\bibitem{m2trans}
Y.~Miura, Y.~Zhang, E.~B. Tsai, C.~P. Langlotz, and D.~Jurafsky, ``Improving factual completeness and consistency of image-to-text radiology report generation,'' {\em arXiv preprint arXiv:2010.10042}, 2020.

\bibitem{rgrg}
T.~Tanida, P.~M{\"u}ller, G.~Kaissis, and D.~Rueckert, ``Interactive and explainable region-guided radiology report generation,'' in {\em Proceedings of the IEEE/CVF Conference on Computer Vision and Pattern Recognition}, pp.~7433--7442, 2023.

\bibitem{chen-emnlp-2020-r2gen}
Z.~Chen, Y.~Song, T.-H. Chang, and X.~Wan, ``Generating radiology reports via memory-driven transformer,'' in {\em Proceedings of the 2020 Conference on Empirical Methods in Natural Language Processing}, Nov. 2020.

\bibitem{wang2018tienet}
X.~Wang, Y.~Peng, L.~Lu, Z.~Lu, and R.~M. Summers, ``Tienet: Text-image embedding network for common thorax disease classification and reporting in chest x-rays,'' in {\em Proceedings of the IEEE conference on computer vision and pattern recognition}, pp.~9049--9058, 2018.

\bibitem{smit2020chexbert}
A.~Smit, S.~Jain, P.~Rajpurkar, A.~Pareek, A.~Y. Ng, and M.~P. Lungren, ``Chexbert: combining automatic labelers and expert annotations for accurate radiology report labeling using bert,'' {\em arXiv preprint arXiv:2004.09167}, 2020.

\bibitem{jain2021radgraph}
S.~Jain, A.~Agrawal, A.~Saporta, S.~Q. Truong, D.~N. Duong, T.~Bui, P.~Chambon, Y.~Zhang, M.~P. Lungren, A.~Y. Ng, {\em et~al.}, ``Radgraph: Extracting clinical entities and relations from radiology reports,'' {\em arXiv preprint arXiv:2106.14463}, 2021.

\bibitem{coco_captions}
X.~Chen, H.~Fang, T.-Y. Lin, R.~Vedantam, S.~Gupta, P.~Doll{\'a}r, and C.~L. Zitnick, ``Microsoft coco captions: Data collection and evaluation server,'' {\em arXiv preprint arXiv:1504.00325}, 2015.

\bibitem{MedFlamingo23}
M.~Moor, Q.~Huang, S.~Wu, M.~Yasunaga, Y.~Dalmia, J.~Leskovec, C.~Zakka, E.~P. Reis, and P.~Rajpurkar, ``Med-flamingo: a multimodal medical few-shot learner,'' in {\em Machine Learning for Health, ML4H@NeurIPS 2023, 10 December 2023, New Orleans, Louisiana, {USA}} (S.~Hegselmann, A.~Parziale, D.~Shanmugam, S.~Tang, M.~N. Asiedu, S.~Chang, T.~Hartvigsen, and H.~Singh, eds.), vol.~225 of {\em Proceedings of Machine Learning Research}, pp.~353--367, {PMLR}, 2023.

\bibitem{fernando2024raddino}
F.~Pérez-García, H.~Sharma, S.~Bond-Taylor, K.~Bouzid, V.~Salvatelli, M.~Ilse, S.~Bannur, D.~C. Castro, A.~Schwaighofer, M.~P. Lungren, M.~Wetscherek, N.~Codella, S.~L. Hyland, J.~Alvarez-Valle, and O.~Oktay, ``Rad-dino: Exploring scalable medical image encoders beyond text supervision,'' 2024.

\bibitem{3dmir2023}
A.~{Ben Abacha}, A.~Santamar{\'{\i}}a{-}Pang, H.~H. Lee, J.~Merkow, Q.~Cai, S.~T. Devarakonda, A.~Islam, J.~Gong, M.~P. Lungren, T.~Lin, N.~C.~F. Codella, and I.~Tarapov, ``3d-mir: {A} benchmark and empirical study on 3d medical image retrieval in radiology,'' {\em CoRR}, vol.~abs/2311.13752, 2023.

\bibitem{BiomedGPT24}
K.~Zhang, R.~Zhou, E.~Adhikarla, Z.~Yan, Y.~Liu, J.~Yu, Z.~Liu, X.~Chen, B.~D. Davison, H.~Ren, {\em et~al.}, ``A generalist vision--language foundation model for diverse biomedical tasks,'' {\em Nature Medicine}, pp.~1--13, 2024.

\bibitem{florence}
L.~Yuan, D.~Chen, Y.-L. Chen, N.~Codella, X.~Dai, J.~Gao, H.~Hu, X.~Huang, B.~Li, C.~Li, {\em et~al.}, ``Florence: A {N}ew {F}oundation {M}odel for {C}omputer {V}ision,'' {\em arXiv preprint arXiv:2111.11432}, 2021.

\bibitem{clip}
A.~Radford, J.~W. Kim, C.~Hallacy, A.~Ramesh, G.~Goh, S.~Agarwal, G.~Sastry, A.~Askell, P.~Mishkin, J.~Clark, G.~Krueger, and I.~Sutskever, ``Learning transferable visual models from natural language supervision,'' 2021.

\bibitem{davit}
M.~Ding, B.~Xiao, N.~Codella, P.~Luo, J.~Wang, and L.~Yuan, ``Da{V}i{T}: {D}ual attention vision transformers,'' in {\em European conference on computer vision}, pp.~74--92, Springer, 2022.

\bibitem{unicl}
J.~Yang, C.~Li, P.~Zhang, B.~Xiao, C.~Liu, L.~Yuan, and J.~Gao, ``Unified {C}ontrastive {L}earning in {I}mage-{T}ext-{L}abel {S}pace,'' 2022.

\bibitem{mimic}
A.~E. Johnson, T.~J. Pollard, S.~J. Berkowitz, N.~R. Greenbaum, M.~P. Lungren, C.-y. Deng, R.~G. Mark, and S.~Horng, ``M{IMIC}-{CXR}, a de-identified publicly available database of chest radiographs with free-text reports,'' {\em Scientific data}, vol.~6, no.~1, p.~317, 2019.

\bibitem{irma2009}
T.~M. Lehmann, H.~Schubert, D.~Keysers, M.~Kohnen, and B.~B. Wein, ``The {IRMA} code for unique classification of medical images,'' in {\em Medical Imaging 2003: PACS and Integrated Medical Information Systems: Design and Evaluation}, vol.~5033, pp.~440--451, SPIE, 2003.

\bibitem{rsnaboneage}
S.~S. Halabi, L.~M. Prevedello, J.~Kalpathy-Cramer, A.~B. Mamonov, A.~Bilbily, M.~Cicero, I.~Pan, L.~A. Pereira, R.~T. Sousa, N.~Abdala, {\em et~al.}, ``The {RSNA} {P}ediatric {B}one {A}ge {M}achine {L}earning {C}hallenge.,'' {\em Radiology}, vol.~290, no.~2, pp.~498--503, 2019.

\bibitem{gpboneage}
C.~M. Gaskin, S.~L. Kahn, J.~C. Bertozzi, and P.~M. Bunch, {\em {Skeletal Development of the Hand and Wrist: A Radiographic Atlas and Digital Bone Age Companion }}.
\newblock Oxford University Press, 11 2011.

\bibitem{upenn}
S.~Bakas, C.~Sako, H.~Akbari, M.~Bilello, A.~Sotiras, G.~Shukla, J.~D. Rudie, N.~F. Santamar{\'\i}a, A.~F. Kazerooni, S.~Pati, {\em et~al.}, ``The {U}niversity of {P}ennsylvania glioblastoma ({UP}enn-{GBM}) cohort: advanced {MRI}, clinical, genomics, \& radiomics,'' {\em Scientific data}, vol.~9, no.~1, p.~453, 2022.

\bibitem{tcgasarc}
C.~Roche, E.~Bonaccio, and J.~Filippini, ``The {C}ancer {G}enome {A}tlas {S}arcoma {C}ollection ({TCGA-SARC}) ({V}ersion 3) [{D}ata set]. {T}he {C}ancer {I}maging {A}rchive.,'' {\em The Cancer Imaging Archive}, 2016.

\bibitem{sd198}
X.~Sun, J.~Yang, M.~Sun, and K.~Wang, ``A {B}enchmark for {A}utomatic {V}isual {C}lassification of {C}linical {S}kin {D}isease {I}mages,'' in {\em Computer Vision--ECCV 2016: 14th European Conference, Amsterdam, The Netherlands, October 11-14, 2016, Proceedings, Part VI 14}, pp.~206--222, Springer, 2016.

\bibitem{isic2017}
N.~C.~F. Codella, D.~Gutman, M.~E. Celebi, B.~Helba, M.~A. Marchetti, S.~W. Dusza, A.~Kalloo, K.~Liopyris, N.~Mishra, H.~Kittler, and A.~Halpern, ``Skin lesion analysis toward melanoma detection: {A} challenge at the 2017 {I}nternational symposium on biomedical imaging ({ISBI}), hosted by the international skin imaging collaboration ({ISIC}),'' in {\em 2018 IEEE 15th International Symposium on Biomedical Imaging (ISBI 2018)}, pp.~168--172, 2018.

\bibitem{isic2018}
N.~C.~F. Codella, V.~Rotemberg, P.~Tschandl, M.~E. Celebi, S.~W. Dusza, D.~A. Gutman, B.~Helba, A.~Kalloo, K.~Liopyris, M.~A. Marchetti, H.~Kittler, and A.~Halpern, ``Skin lesion analysis toward melanoma detection 2018: {A} challenge hosted by the {I}nternational skin imaging collaboration {(ISIC)},'' {\em CoRR}, vol.~abs/1902.03368, 2019.

\bibitem{isic2019}
M.~Combalia, N.~C.~F. Codella, V.~Rotemberg, B.~Helba, V.~Vilaplana, O.~Reiter, C.~Carrera, A.~Barreiro, A.~C. Halpern, S.~Puig, and J.~Malvehy, ``{BCN20000}: {D}ermoscopic lesions in the wild,'' 2019.

\bibitem{pcam}
B.~Ehteshami~Bejnordi, M.~Veta, P.~Johannes~van Diest, B.~van Ginneken, N.~Karssemeijer, G.~Litjens, J.~A. W.~M. van~der Laak, , and the CAMELYON16~Consortium, ``{Diagnostic {A}ssessment of {D}eep {L}earning {A}lgorithms for {D}etection of {L}ymph {N}ode {M}etastases in {W}omen With {B}reast {C}ancer},'' {\em JAMA}, vol.~318, pp.~2199--2210, 12 2017.

\bibitem{rsnamammography}
C.~Carr, F.~Kitamura, J.~Kalpathy-Cramer, J.~Mongan, K.~Andriole, M.~Vazirabad, M.~Riopel, R.~Ball, and S.~Dane, ``{RSNA} {S}creening {M}ammography {B}reast {C}ancer {D}etection. 2022.''

\bibitem{lidcidri}
K.~S. Mader, ``The {L}ung {I}mage {D}atabase {C}onsortium image collection ({LIDC-IDRI}),'' {\em IEEE Dataport}, 2021.

\bibitem{padufes20}
A.~G. Pacheco, G.~R. Lima, A.~S. Salomao, B.~Krohling, I.~P. Biral, G.~G. de~Angelo, F.~C. Alves~Jr, J.~G. Esgario, A.~C. Simora, P.~B. Castro, {\em et~al.}, ``P{AD-UFES}-20: A skin lesion dataset composed of patient data and clinical images collected from smartphones,'' {\em Data in brief}, vol.~32, p.~106221, 2020.

\bibitem{odir5k}
G.~Challenge, ``Peking {U}niversity {I}nternational {C}ompetition on {O}cular {D}isease {I}ntelligent {R}ecognition (odir-2019),'' 2019.

\bibitem{oct2017}
D.~Kermany, ``Labeled {O}ptical {C}oherence {T}omography ({OCT}) for {C}lassification,'' 2017.

\bibitem{busi}
W.~Al-Dhabyani, M.~Gomaa, H.~Khaled, and A.~Fahmy, ``Dataset of breast ultrasound images,'' {\em Data in brief}, vol.~28, p.~104863, 2020.

\bibitem{hmcqu}
T.~U. Hamad Medical Corporation Heart~Hospital, Qatar~University, ``{HMC-QU} {D}ataset,'' 2021.

\bibitem{msd}
M.~A. et~al., ``The {M}edical {S}egmentation {D}ecathlon,'' {\em Nature Communications}, vol.~13, 2022.

\bibitem{roco}
O.~Pelka, S.~Koitka, J.~R{\"u}ckert, F.~Nensa, and C.~M. Friedrich, ``Radiology {O}bjects in {C}ontext ({ROCO}): {A} {M}ultimodal {I}mage {D}ataset,'' in {\em Intravascular Imaging and Computer Assisted Stenting and Large-Scale Annotation of Biomedical Data and Expert Label Synthesis: 7th Joint International Workshop, CVII-STENT 2018 and Third International Workshop, LABELS 2018, Held in Conjunction with MICCAI 2018, Granada, Spain, September 16, 2018, Proceedings 3}, pp.~180--189, Springer, 2018.

\bibitem{oct2017b}
D.~Kermany, K.~Zhang, and M.~Goldbaum, ``Large {D}ataset of {L}abeled {O}ptical {C}oherence {T}omography {(OCT)} and {C}hest {X}-{R}ay {I}mages,'' 2018.

\end{thebibliography}


\section{Appendix}

\subsection{Medical Datasets}
\label{sec:appendixdatasets}


The following itemization describes all the public and private datasets leveraged to train and evaluate this model (summarized in Table~\ref{tab:pretraining}). For each dataset, images were associated with text descriptions constituted from either sections of associated clinical reports, DICOM fields, or in the case of classification datasets, the imaging modality, body region, view (where applicable), diagnosis, and any other descriptive text that may help differentiate the category from other categories. Images were standardized to dimensions of 512x512, not respecting aspect ratios, encoded as JPEG base64 strings (mix of 75\% and 98\% quality). DICOMs were converted to 8-bit gray-scale images using standard window levels in each imaging setting. In total, 14 unique domains were pooled to a unified dataset with approximately 1,000 descriptive text categorical labels, and several hundred thousands of text descriptions \footnote{See GitHub for a list of all of the descriptive text labels that we used}. The 14 domains include Chest X-Ray Diagnostics, X-Ray Exam Parameters, Bone Age, Mammography, Chest CT Diagnostics, MR Exam Parameters, CT Exam Parameters, Breast Ultrasound, Echocardiography Exam Parameter, Dermoscopy Diagnostics, Clinical Photography Diagnostics, Histopathology Diagnostics, OCT Diagnostics, and Fundus Imaging Diagnostics. 

\begin{itemize}
    \item {\bf MIMIC-CXR ~\cite{mimic}:} Frontal chest X-rays from the training partition of the MIMIC-CXR dataset and the associated text reports. Rule-based processing was carried out to extract findings and impressions separately, or to map non-labeled report sections to the relevant sections. During training, text is randomly sampled from either the findings or the impression section. In total 203,170 images from this dataset were used for training. 
    \item {\bf NIH-CXR-LT ~\cite{ltcxr}:} The NIH-CXR-LT dataset contains long tail distribution categories spanning 20 disease classes for frontal chest X-rays. 68,058 images from the training dataset were leveraged for training. For evaluation, 20,279 images from the established test set was used.
    \item {\bf IRMA 2009 ~\cite{irma2009}:} A dataset containing X-rays covering a spectrum of body regions, views, and patient positions. Category information is specified in a coding system, with a PDF mapping the coding system to text for each of the code sub-parts. We converted the coding scheme to the text counterparts by extracting this mapping from the PDF, and leveraged the image and code-text pairs for training. In total, 12,677 image-text pairs were used for training, and 1,733 were held out for evaluation. 
    \item {\bf RSNA BoneAge ~\cite{rsnaboneage}:} Pediatric bone-age hand X-rays annotated with the development age of the images. The images are supplied in 8-bit format with inconsistent window leveling. Preprocessing was applied including histogram equalization followed by window leveling to control and standardize the appearance of the images for subsequent training and inference.  The development age and gender of the image was converted to text using a standardized template. 12,611 images from the training partition are leveraged for pre-training, and evaluations are performed on the established 1,425 validation partition. 
    \item {\bf Gruelich and Pyle (GP) BoneAge Atlas ~\cite{gpboneage}:} Pediatric bone-age hand X-rays annotated with the development age of the images. Sourced 59 images from the original atlas for pretraining. 
    \item {\bf UPENN ~\cite{upenn}:} A dataset of MRI images of glioblastomas. Images were paired with the text of their DICOM image series descriptions. In total 4,645 images with associated texts were organized for training.
    \item {\bf TCGA ~\cite{tcgasarc}}: A multi-modal dataset of imaging for sarcoma diagnostics. CT and MRI images were extracted and associated with the text of their series description, constituting 5,643 image and text pairs. 
    \item {\bf SD198 ~\cite{sd198}:} A dataset of clinical photographs of 198 skin lesions crawled from the web. Train and test splits were not made available but based on random 50\% sampling, which we followed for consistency, yielding 3,253 images for training, and 3,331 for evaluation. 
    \item {\bf ISIC2019 ~\cite{isic2017,isic2018,isic2019}:} A collection of dermascopic images of skin lesions, associated with 8 diagnostic states spanning metastatic and non-metastatic disease. 20,268 images from the training partition were leveraged for pretraining, and a random selection of 5,063 images were held out for evaluation. 
    \item {\bf PatchCamelyon ~\cite{pcam}:} Histopathological images of breast tissue depicting the presence or absence of cancer. 262,144 images and associated text labels were used in training. 32,768 images from the established validation partition were used for evaluation. For KNN classification, a subset of 1,000 randomly selected images from the training set were used as the index.
    \item {\bf RSNA Mammography ~\cite{rsnamammography}:} Images from RSNA hosted and managed challenge on breast cancer detection from mammography. The dataset comprises several styles of mammograms with varying window levels and contrasts. No attempt was made to standardize or normalize the images. In total, 43,764 mammograms were leveraged for training, and a random selection of 10,942 for evaluation. 
    \item {\bf LIDC-IDRI ~\cite{lidcidri}:} A dataset of chest CTs depicting lung nodules at various stages of development. Dataset was broken into tiles of 5x5 across images, with tiles labeled for the maturity of lung nodule present in the tile. 80,201 tiles were sampled for training, and 4,574 tiles were sampled for evaluation. 
    \item {\bf PAD-UFES-20 ~\cite{padufes20}}: A collection of clinical photographs of skin lesions taken from mobile devices, where the images have been cropped over the lesion of interest. 6 diseases are represented. According to precedent 2,065 images (90\%) were leveraged for training, and 233 (10\%) for testing.
    \item {\bf ODIR-5K ~\cite{odir5k}:} Fundus images, where pairs of eyes were annotated across 6 categories. If one eye is not normal, the pair is labeled with the disease of the abnormal eye. Laterality specific textual descriptions were also available. Upon further processing, we discovered about 79 unique textual descriptions were assigned across 6,495 unique eyes, and opted to use these descriptions as labels instead of the reduced 6 labels. 5228 images were used for training, and 1267 images were used for evaluation, which constituted a random 20\% sampling of the top 30 categories (with 10 or more instances in the dataset).
    \item {\bf OCT2017 ~\cite{oct2017}:} The OCT2017 dataset comprises Optical Coherence Tomography (OCT) images, which are a form of microscopic-level images of cross-sectional slices of the retina. From the OCT2017 dataset, we leveraged 83,484 images for training, and 968 images for evaluation. The dataset covered 4 categories of choroidal neovascularization, diabetic macular edema, drusen, and normal retina.
    \item {\bf OCT2018 ~\cite{oct2017b}:} The OCT2018 dataset is an expansion of the OCT2017 dataset, with additional training data and 1,000 test instances comprised of new images. Only the new test dataset was leveraged for evaluation purposes and to faciliate direct comparison with BiomedGPT ~\cite{biomedgpt}. 
    \item {\bf Breast Ultrasound Image (BUSI) ~\cite{busi}:} The BUSI dataset is a small dataset of breast ultrasound images for the diagnosis of breast cancer. 623 images are available for training, and 157 for evaluation, over 3 categories of normal tissue, benign lesion, and malignant lesion. 
    \item {\bf HMC-QU ~\cite{hmcqu}:} A collection of cardiac echocardiographic videos across two views (2 and 4 chamber views) depicting various states of wall-motion abnormalities for the detection of myocardial infarction. For the purposes of our work, the data was used for echo view determination (2 vs 4 chamber view). 4,714 frames were extracted for pre-training, and 2,132 for evaluation. Care was taken to split according to patient exams and not by frame to prevent the same exams occurring in both train and test.
    \item {\bf Medical Segmentation Decathlon (MSD) ~\cite{msd}:} The MSD dataset focuses on CT images of four key organs of the chest and abdomen: the liver, colon, pancreas, and lung. We converted this dataset to image-text pairs for pretraining. To do so, we leveraged TotalSegmentator (v1.5.7), which creates a segmentation map for 104 organs. From the 3D segmentation mask, we quantify the area of each organ in the corresponding 2D slice. This quantification is used to generate captions for the organs present in each 2D slice. The text then comprises modality, body region (chest/abdomen), a list of organs, as well as the presence or absence of metastatic disease. To convert 16-bit DICOM images to 8-bit monochromatic PNG images, we selected a dynamic range of -1000 HU to 1000 HU (Hounsfield Units). This range represents the typical values found in CT images. The dynamic range is then linearly transformed to [0, 255], corresponding to 8-bit depth. 116,500 image-text pairs in total were created for pretraining.
    \item {\bf Bing Echo Few-Shot:} Bing Image Search was crawled for 7 categories representing unique echocardiography views, including 2-chamber, 4-chamber, 5-chamber, short-axis, parasternal long axis view, M-mode doppler, and CW-doppler. Results were manually curated by medical imaging scientists to improve annotation quality. 100 images were curated for training, and 54 for evaluation.
    \item {\bf Bing Exam Parameter Few-Shot:} Bing Image Search was crawled for 21 categories reprenting mixes of modalities (CT, MRI), body regions (Abdomen, Chest, Brain, Knee, C-Spine, L-Spine), and views (Axial, Coronal, Saggital). Results were manually curated by medical imaging scientists to improve annotation quality. 1202 images were used in pretraining, and 401 were held-out for evaluation. 
    \item {\bf Radiological Objects in COntext (ROCO) ~\cite{roco}:} The ROCO dataset is a collection of medical images and associated caption texts. 70,308 images from the ROCO dataset was leveraged for pretraining.
    \item {\bf PMC-15M~\cite{PMC15M}:} We leverage a subset of about 1 million (952,556) sub-figures and associated sub-figure captions from the PMC-15M dataset for pretraining, which were sourced from PubMed Central (PMC). Graphs and charts were filtered out prior to random selection. 
    \item{\bf Partner Datasets (MGB):} We collaborated with Mass General Brigham (MGB) hospital to collect two real world datasets: 1,415 images of chest x-rays annotated for 80 conditions that represent the most clinically significant finding in the image, and 120 pediatric boneage hand x-rays. Of the chest x-rays, 731 were leveraged for pretraining, and 684 for evaluation (referred to as ``MGB CXR''). Of the boneage images, all 120 were leveraged for evaluation only (referred to as ``MGB BoneAge''). 
    \item {\bf Proprietary Datasets:} Multiple other proprietary datasets, composed of procured data, data supplied by collaborative partners, internally annotated data, and data crawled from the web were additionally leveraged for training.  
\end{itemize}

\subsection{Independent Clinical Evaluation: Fairness Assessment}
\label{sec:clinicalfairness}

Table ~\ref{tab:fairness_gender} shows the results of our fairness assessment with respect to gender labels. Table ~\ref{tab:fairness_age} shows the results of our fairness assessment with respect to age.

\begin{table}[h!]
  \centering
  \setlength{\extrarowheight}{1.0pt}  
  \small
  \begin{threeparttable}
    \caption{Gender fairness assessment for AUC results from independent clinical evaluation using KNN classification from extracted image features}
    \label{tab:fairness_gender}
    \setlength{\tabcolsep}{1.55mm}{    
        \begin{tabularx}{\textwidth}{llccccccccc}
            \toprule
            \multirow{3}{*}{Model} & \multirow{3}{*}{Gender} & \multicolumn{8}{c}{Multi-class } & \multicolumn{1}{c}{Binary-class} \\ \cmidrule(r){3-10} \cmidrule(l){11-11} %
            & & ETT &  NGT &  PP & PT & RFrac &  VLine &  Normal & mAUC & Abnormal  \\ \midrule 
            Med-Flamingo & F & 64.5 & 74.7 & 73.2 & 70.2 & 87.7 & 66.4 & 85.4 & 74.6 & 85.9  \\
            & M & 68.7 & 69.0 & 62.4 & 68.1 & 85.3 & 67.6 & 81.3 & 71.8 & 81.1  \\
            \midrule

            BiomedCLIP & F & 79.1 & 83.2 & 81.1 & 76.1 & 90.9 & 75.1 & 90.9 & 82.3 & 88.5  \\
            & M & 79.0 & 78.1 & 69.7 & 75.1 & 88.9 & 69.2 & 89.7 & 78.5 & 88.6  \\
            \midrule

            RAD-DINO & F & 83.6 & 84.7 & 82.3 & 77.3 & 91.1 & 83.2 & 91.4 & 84.8 & 91.4  \\
            & M & 82.0 & 81.1 & 79.5 & 78.4 & 91.8 & 79.7 & 89.5 & 83.1 & 89.4 \\
            \midrule

            \ourmodel & F & 91.1 & 90.5 & 91.1 & 92.2 & 95.4 & 90.5 & 96.9 & 92.5 & 96.8  \\
            & M & 84.9 & 89.0 & 88.9 & 90.9 & 94.8 & 87.5 & 96.6 & 90.4 & 96.5  \\
            \bottomrule
        \end{tabularx}
    }
    \begin{tablenotes}
      \footnotesize
      \item[*] ETT=endotracheal tube, NGT=nasogastric tube, PP=Pneumoperitoneum, PT=pneumothorax, RFrac=rib fracture, VLine=vascular lines, F=female patients (709 in total), M=male patients (1059 in total).
    \end{tablenotes}
  \end{threeparttable}
\end{table}

\begin{table}[h!]
  \centering
  \setlength{\extrarowheight}{1.0pt}  
  \small
  \begin{threeparttable}
    \caption{Age fairness assessment for AUC results from independent clinical evaluation using KNN classification from extracted image features}
    \label{tab:fairness_age}
    \setlength{\tabcolsep}{1.4mm}{    
        \begin{tabularx}{\textwidth}{llccccccccc}
\toprule
            \multirow{3}{*}{Model} & \multirow{3}{*}{Age} & \multicolumn{8}{c}{Multi-class } & \multicolumn{1}{c}{Binary-class} \\ \cmidrule(r){3-10} \cmidrule(l){11-11} %
            & & ETT &  NGT &  PP & PT & RFrac &  VLine &  Normal & mAUC & Abnormal  \\ \midrule 
            \multirow{5}{*}{Med-Flamingo} & $\leq 20$ & 100.0 & 67.9 & 98.6 & 78.4 & 97.1 & 57.2 & 99.4 & 85.5 & 98.8  \\
            & $\leq 40$ & 72.3 & 74.4 & 61.5 & 73.9 & 86.9 & 67.7 & 86.8 & 74.8 & 86.9  \\
            & $\leq 60$ & 67.2 & 69.0 & 68.0 & 63.9 & 89.0 & 70.7 & 86.6 & 73.5 & 86.9  \\
            & $\leq 80$ & 63.7 & 69.2 & 58.3 & 66.6 & 87.4 & 65.0 & 76.2 & 69.5 & 75.3  \\
            & $\leq 100$ & 69.3 & 59.2 & 80.7 & 72.6 & 81.3 & 58.3 & 82.9 & 72.0 & 82.5  \\
            \midrule

            \multirow{5}{*}{BiomedCLIP} & $\leq 20$ & 96.9 & 75.9 & 87.0 & 84.0 & 100.0 & 61.5 & 98.8 & 86.3 & 98.2  \\
            & $\leq 40$ & 79.4 & 86.5 & 81.1 & 79.1 & 92.7 & 77.4 & 88.9 & 83.6 & 88.6  \\
            & $\leq 60$ & 79.0 & 80.0 & 74.3 & 65.1 & 91.6 & 73.9 & 92.2 & 79.4 & 90.6  \\
            & $\leq 80$ & 77.0 & 74.5 & 70.0 & 77.4 & 90.5 & 68.2 & 88.3 & 78.0 & 86.1  \\
            & $\leq 100$ & 88.5 & 74.3 & 69.2 & 68.8 & 80.9 & 72.3 & 83.0 & 76.7 & 77.8  \\
            \midrule

            \multirow{5}{*}{RAD-DINO} & $\leq 20$ & 100.0 & 76.8 & 80.4 & 83.5 & 89.6 & 72.4 & 99.7 & 86.1 & 99.7  \\
            & $\leq 40$ & 79.6 & 82.1 & 79.2 & 75.9 & 94.4 & 83.1 & 92.2 & 83.8 & 91.9  \\
            & $\leq 60$ & 80.8 & 84.4 & 81.8 & 79.0 & 92.9 & 83.8 & 91.7 & 84.9 & 92.0  \\
            & $\leq 80$ & 81.4 & 79.5 & 84.4 & 77.5 & 93.6 & 78.0 & 87.8 & 83.2 & 87.0  \\
            & $\leq 100$ & 88.1 & 77.2 & 57.5 & 77.1 & 83.8 & 80.6 & 90.2 & 79.2 & 90.6  \\
            \midrule

            \multirow{5}{*}{\ourmodel} & $\leq 20$ & 98.8 & 83.2 & 100.0 & 99.4 & 100.0 & 77.8 & 99.7 & 94.1 & 99.7  \\
            & $\leq 40$ & 88.3 & 89.4 & 79.9 & 87.5 & 97.2 & 88.9 & 97.5 & 89.8 & 97.6  \\
            & $\leq 60$ & 83.8 & 90.8 & 88.9 & 90.6 & 95.0 & 90.1 & 97.7 & 91.0 & 97.7  \\
            & $\leq 80$ & 86.8 & 88.2 & 91.1 & 91.3 & 95.7 & 89.4 & 93.7 & 90.9 & 93.6  \\
            & $\leq 100$ & 94.7 & 88.6 & 90.1 & 91.0 & 94.3 & 86.8 & 99.0 & 92.1 & 98.6  \\
            \bottomrule
        \end{tabularx}
    }
    \begin{tablenotes}
      \footnotesize
        \item[*] ETT = endotracheal tube, NGT = nasogastric tube, PP = Pneumoperitoneum, PT = pneumothorax, RFrac = rib fracture, VLine = vascular lines. A total of 164 patients with age $\leq 20$, 237 patients with age $\leq 40$, 570 patients with age $\leq 60$, 610 patients with age $\leq 80$, and 99 patients with age $\leq 100$ are involved in the analysis. Patients older than 100 are unidentified and are excluded from analysis.
    \end{tablenotes}
  \end{threeparttable}
\end{table}

\subsection{Report Generation Performance}
\label{sec:reportgen}

\begin{figure}[h!]
    \centering
    \includegraphics[width=1.0\textwidth]{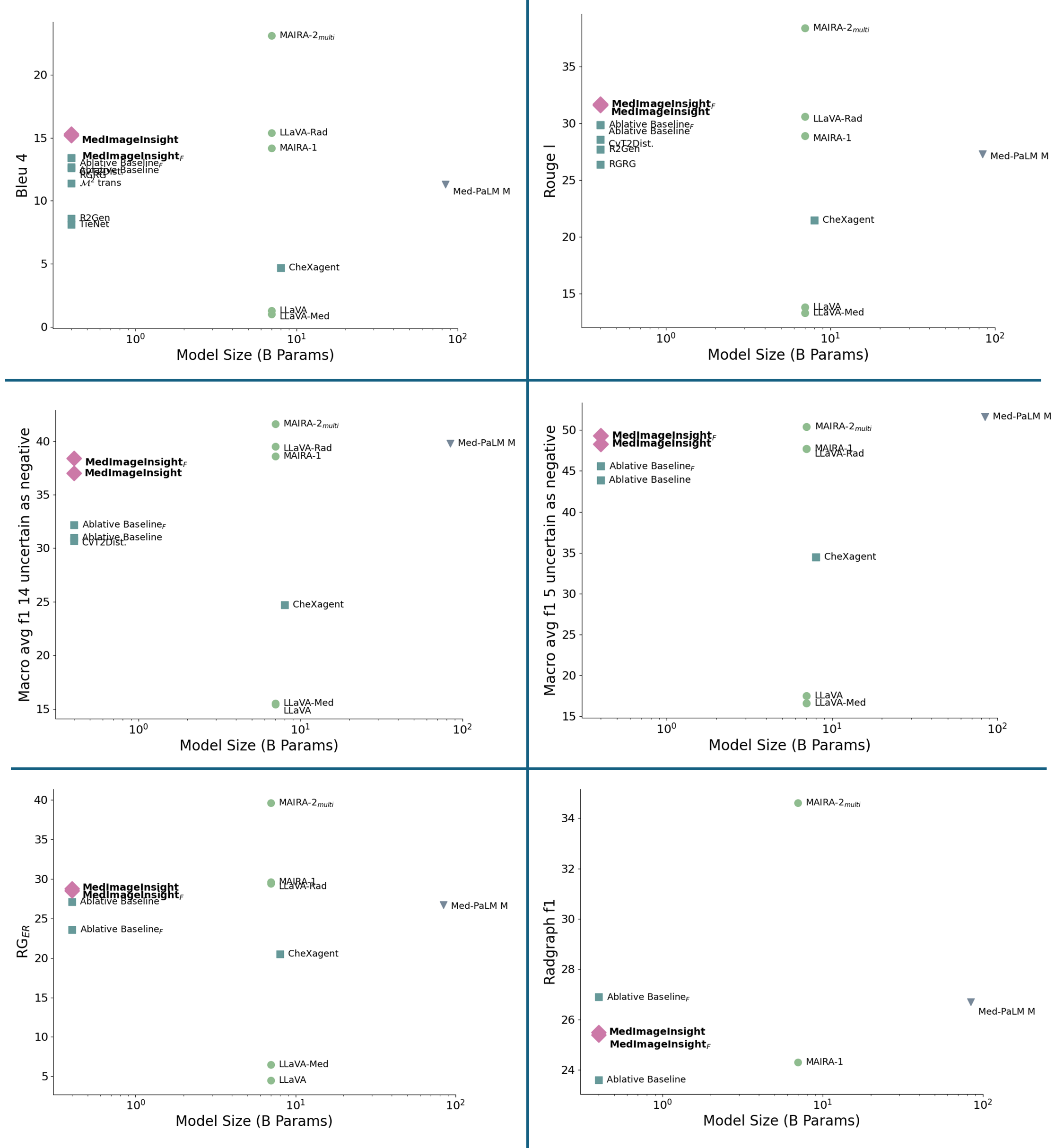}
    \caption{MIMIC-CXR metrics according to performance and model size. Multi-image MIMIC-CXR benchmarks denoted with subscript ``multi''.}
    \label{fig:reportgentables}
\end{figure}

The following section includes a graph of report generation performance according to multiple metrics, plotted against model size (Fig. ~\ref{fig:reportgentables}).

\subsection{Category-level Performance}
\label{sec:categoryperf}

The following section describes the category-level performance across datasets, as measured by Area Under Curve (AUC). 

\begin{table}[h!]
  \centering
  \begin{threeparttable}
    \caption{\ourmodel LTCXR Dataset Category Performance}
    \label{tab:ltcxrcp}
    \begin{tabular}{@{}lc@{}}
      \toprule
      Category & AUC  \\
      \midrule 
        x-ray chest anteroposterior Atelectasis & 0.825\\
        x-ray chest anteroposterior Calcification of the Aorta & 0.937\\
        x-ray chest anteroposterior Cardiomegaly & 0.927\\
        x-ray chest anteroposterior Consolidation & 0.788\\
        x-ray chest anteroposterior Edema & 0.913\\
        x-ray chest anteroposterior Effusion & 0.864\\
        x-ray chest anteroposterior Emphysema & 0.866\\
        x-ray chest anteroposterior Fibrosis & 0.811\\
        x-ray chest anteroposterior Hernia & 0.961\\
        x-ray chest anteroposterior Infiltration & 0.681\\
        x-ray chest anteroposterior Mass & 0.867\\
        x-ray chest anteroposterior Nodule & 0.831\\
        x-ray chest anteroposterior No Finding & 0.554\\
        x-ray chest anteroposterior Pleural Thickening & 0.793\\
        x-ray chest anteroposterior Pneumomediastinum & 0.926\\
        x-ray chest anteroposterior Pneumonia & 0.731\\
        x-ray chest anteroposterior Pneumoperitoneum & 0.954\\
        x-ray chest anteroposterior Pneumothorax & 0.906\\
        x-ray chest anteroposterior Subcutaneous Emphysema & 0.963\\
        x-ray chest anteroposterior Tortuous Aorta & 0.962\\

      \bottomrule
    \end{tabular}
  \end{threeparttable}
\end{table}

\begin{table}[h!]
  \centering
  \begin{threeparttable}
    \caption{\ourmodel IRMA Dataset Category Performance (1/3)}
    \label{tab:irmacp}
    \begin{tabular}{lc} 
      \toprule
      Category & AUC  \\
      \midrule 
x-ray abdomen lower abdomen lower middle quadrant, coronal AP  supine & 0.941\\
x-ray abdomen lower abdomen lower right quadrant, coronal PA  supine & 0.991\\
x-ray abdomen middle abdomen middle right abdomen, coronal AP & 0.995\\
x-ray abdomen middle abdomen peri navel region, coronal PA  supine & 0.995\\
x-ray abdomen middle abdomen, coronal AP  supine & 0.986\\
x-ray abdomen middle abdomen, coronal PA  supine & 0.980\\
x-ray abdomen, coronal AP  lateral decubitus & 0.991\\
x-ray abdomen, coronal AP  supine & 0.976\\
x-ray abdomen, coronal PA  lateral decubitus & 0.969\\
x-ray abdomen, coronal PA  upright & 0.995\\
x-ray abdomen upper abdomen, coronal PA  upright & 0.993\\
x-ray abdomen upper abdomen upper middle quadrant, coronal AP & 0.890\\
x-ray breast , axial craniocaudal & 1.000\\
x-ray breast , other orientation oblique & 0.997\\
x-ray chest bones lower ribs, coronal AP & 0.992\\
x-ray chest bones lower ribs, other orientation left anterior oblique  & 0.996\\
x-ray chest bones, coronal AP & 1.000\\
x-ray chest bones upper ribs, coronal AP & 0.985\\
x-ray chest bones upper ribs, other orientation right anterior oblique  & 0.995\\
x-ray chest mediastinum, coronal PA & 0.893\\
x-ray chest, coronal AP  inspiration & 0.993\\
x-ray chest, coronal AP  supine & 0.957\\
x-ray chest, coronal PA  expiration & 1.000\\
x-ray chest, coronal PA & 0.963\\
x-ray chest, sagittal lateral, right-left inspiration & 0.995\\
x-ray cranium cranial base petrous bone, sagittal lateral, left-right & 0.965\\
x-ray cranium cranial base petrous bone, sagittal lateral, right-left & 0.999\\
x-ray cranium facial cranium eye area, other orientation occipitofrontal & 0.998\\
x-ray cranium facial cranium eye area, sagittal lateral, left-right & 0.996\\
x-ray cranium facial cranium eye area, sagittal lateral, right-left & 0.994\\
x-ray cranium facial cranium mandible, other orientation occipitomental & 1.000\\
x-ray cranium facial cranium mandible, sagittal lateral, left-right & 0.991\\
x-ray cranium facial cranium nose area, coronal PA & 0.993\\
x-ray cranium facial cranium nose area, other orientation occipitofrontal & 0.999\\
x-ray cranium facial cranium nose area, other orientation occipitomental & 0.999\\
x-ray cranium facial cranium nose area, sagittal lateral, left-right & 0.995\\
x-ray cranium facial cranium nose area, sagittal lateral, right-left & 0.997\\
x-ray cranium facial cranium temporo mandibular area, other orientation & 1.000\\
x-ray cranium neuro cranium occipital area, other orientation & 1.000\\
x-ray cranium neuro cranium, sagittal lateral, left-right & 0.999\\
x-ray cranium neuro cranium, sagittal lateral, right-left & 1.000\\
x-ray cranium, coronal AP & 1.000\\

      \bottomrule
    \end{tabular}
  \end{threeparttable}
\end{table}

\begin{table}[h!]
  \centering
  \begin{threeparttable}
    \caption{\ourmodel IRMA Dataset Category Performance (2/3)}
    \label{tab:irmacpb}
    \begin{tabular}{lc}
      \toprule
      Category & AUC  \\
      \midrule 
x-ray lower extremity : leg ankle joint, coronal AP & 0.997\\
x-ray lower extremity : leg ankle joint, sagittal mediolateral & 0.997\\
x-ray lower extremity : leg foot middle foot, coronal AP & 0.985\\
x-ray lower extremity : leg foot tarsal bones, sagittal mediolateral & 0.993\\
x-ray lower extremity : leg foot toe, coronal AP & 0.999\\
x-ray lower extremity : leg foot toe, sagittal mediolateral & 0.998\\
x-ray lower extremity : leg foot, coronal AP & 0.995\\
x-ray lower extremity : leg foot, coronal PA  sitting & 1.000\\
x-ray lower extremity : leg foot, other orientation left anterior oblique  & 0.993\\
x-ray lower extremity : leg foot, other orientation right anterior oblique  & 0.995\\
x-ray lower extremity : leg foot, sagittal mediolateral & 0.996\\
x-ray lower extremity : leg hip, coronal AP & 0.995\\
x-ray lower extremity : leg hip, sagittal mediolateral & 0.996\\
x-ray lower extremity : leg knee patella, axial caudocranial  & 1.000\\
x-ray lower extremity : leg knee, coronal AP & 0.989\\
x-ray lower extremity : leg knee, sagittal mediolateral & 0.995\\
x-ray lower extremity : leg lower leg, coronal AP & 0.988\\
x-ray lower extremity : leg lower leg, sagittal mediolateral & 0.983\\
x-ray lower extremity : leg, coronal AP & 0.991\\
x-ray lower extremity : leg, sagittal mediolateral & 0.996\\
x-ray lower extremity : leg upper leg distal upper leg, coronal AP & 0.998\\
x-ray lower extremity : leg upper leg proximal upper leg, sagittal mediolateral & 0.943\\
x-ray lower extremity : leg upper leg, coronal AP & 0.992\\
x-ray lower extremity : leg upper leg, sagittal mediolateral & 0.992\\
x-ray pelvis iliac bone, coronal AP & 0.987\\
x-ray pelvis sarcral bone iliosacral-junction, coronal AP & 0.992\\
x-ray pelvis sarcral bone, sagittal & 0.994\\
x-ray pelvis, coronal AP  supine & 0.985\\
x-ray pelvis, coronal AP & 0.994\\
x-ray spine cervical spine dens, coronal AP & 1.000\\
x-ray spine cervical spine, coronal AP & 0.998\\
x-ray spine cervical spine, other orientation right anterior oblique  & 0.998\\
x-ray spine cervical spine, sagittal lateral, left-right & 0.993\\
x-ray spine cervical spine, sagittal lateral, right-left inclination & 0.982\\
x-ray spine cervical spine, sagittal lateral, right-left reclination & 0.985\\
x-ray spine cervical spine, sagittal lateral, right-left & 0.999\\
x-ray spine cervical spine upper cervical spine, coronal AP & 1.000\\
x-ray spine cervical spine upper cervical spine, sagittal lateral, right-left & 0.994\\
x-ray spine lumbar spine lower lumbar spine, sagittal lateral, right-left & 0.991\\
x-ray spine lumbar spine thoraco-lumbar conjunction, coronal AP & 0.987\\
x-ray spine lumbar spine thoraco-lumbar conjunction, sagittal lateral, right-left & 0.995\\
x-ray spine lumbar spine, coronal AP & 0.997\\
x-ray spine lumbar spine, sagittal lateral, right-left & 0.997\\
x-ray spine sacral bone, sagittal lateral, right-left & 0.993\\
x-ray spine thoracic spine cervico-thoracic conjunction, sagittal lateral, right-left & 0.996\\
x-ray spine thoracic spine, coronal AP & 0.998\\
x-ray spine thoracic spine, sagittal lateral, left-right & 0.970\\
x-ray spine thoracic spine, sagittal lateral, right-left & 0.994\\
x-ray spine, sagittal & 0.990\\

      \bottomrule
    \end{tabular}

  \end{threeparttable}
\end{table}

\begin{table}[h!]
  \centering
  \begin{threeparttable}
    \caption{\ourmodel IRMA Dataset Category Performance (3/3)}
    \label{tab:irmacpc}
    \begin{tabular}{lc}
      \toprule
      Category & AUC  \\
      \midrule 
x-ray upper extremity : arm elbow, coronal AP & 0.998\\
x-ray upper extremity : arm elbow, sagittal lateromedial & 0.999\\
x-ray upper extremity : arm forearm, coronal AP & 0.997\\
x-ray upper extremity : arm forearm, sagittal lateromedial & 0.994\\
x-ray upper extremity : arm hand carpal bones, coronal AP & 0.996\\
x-ray upper extremity : arm hand carpal bones, other orientation right anterior oblique  & 0.980\\
x-ray upper extremity : arm hand carpal bones, sagittal lateromedial & 0.998\\
x-ray upper extremity : arm hand finger, coronal PA & 0.998\\
x-ray upper extremity : arm hand finger, sagittal & 0.999\\
x-ray upper extremity : arm hand middle hand, coronal AP & 0.991\\
x-ray upper extremity : arm hand middle hand, other orientation left anterior oblique  & 0.990\\
x-ray upper extremity : arm hand middle hand, other orientation right anterior oblique  & 0.993\\
x-ray upper extremity : arm hand middle hand, sagittal & 0.985\\
x-ray upper extremity : arm hand, coronal PA & 0.998\\
x-ray upper extremity : arm hand, other orientation left anterior oblique  & 0.995\\
x-ray upper extremity : arm hand, other orientation right anterior oblique  & 0.992\\
x-ray upper extremity : arm hand, sagittal mediolateral & 0.983\\
x-ray upper extremity : arm radio carpal joint, coronal AP & 0.991\\
x-ray upper extremity : arm radio carpal joint, sagittal lateromedial & 0.992\\
x-ray upper extremity : arm shoulder acromio-scapula joint, coronal AP & 0.992\\
x-ray upper extremity : arm shoulder acromio-scapula joint, sagittal mediolateral & 0.997\\
x-ray upper extremity : arm shoulder humero-scapular joint, coronal AP & 0.993\\
x-ray upper extremity : arm shoulder humero-scapular joint, sagittal mediolateral & 0.998\\
x-ray upper extremity : arm shoulder, coronal AP & 0.961\\
x-ray upper extremity : arm, coronal AP & 0.997\\
x-ray upper extremity : arm upper arm distal upper arm, coronal AP & 0.972\\
x-ray upper extremity : arm upper arm distal upper arm, sagittal mediolateral & 0.984\\
x-ray upper extremity : arm upper arm proximal upper arm, coronal AP & 0.989\\
x-ray upper extremity : arm upper arm proximal upper arm, sagittal mediolateral & 0.995\\
x-ray upper extremity : arm upper arm, coronal AP & 0.986\\
x-ray upper extremity : arm upper arm, sagittal lateral, left-right & 0.992\\
x-ray upper extremity : arm upper arm, sagittal lateral, right-left & 0.995\\
x-ray upper extremity : arm upper arm, sagittal mediolateral & 0.955\\

      \bottomrule
    \end{tabular}

  \end{threeparttable}
\end{table}

\begin{table}[h!]
  \centering
  \begin{threeparttable}
    \caption{\ourmodel MGB CXR Dataset Category Performance (1/2)}
    \label{tab:mgbcxr}
    \begin{tabular}{@{}lc@{}}
      \toprule
      Category & AUC  \\
      \midrule 
x-ray chest anteroposterior acute clavicle fracture, priority diagnosis & 0.985 \\
x-ray chest anteroposterior acute humerus fracture, priority diagnosis & 0.997 \\
x-ray chest anteroposterior acute rib fracture, priority diagnosis & 0.766 \\
x-ray chest anteroposterior atelectasis, priority diagnosis & 0.830 \\
x-ray chest anteroposterior bronchiectasis, priority diagnosis & 0.960 \\
x-ray chest anteroposterior calcified pleural plaques, priority diagnosis & 0.998 \\
x-ray chest anteroposterior cardiac valve prosthesis, priority diagnosis & 0.940 \\
x-ray chest anteroposterior cavitating mass with content, priority diagnosis & 0.984 \\
x-ray chest anteroposterior chronic rib fracture, priority diagnosis & 0.996 \\
x-ray chest anteroposterior diaphragmatic elevation, priority diagnosis & 0.990 \\
x-ray chest anteroposterior diaphragmatic eventration, priority diagnosis & 0.989 \\
x-ray chest anteroposterior diffuse airspace opacities, priority diagnosis & 0.964 \\
x-ray chest anteroposterior diffuse bullae, priority diagnosis & 0.984 \\
x-ray chest anteroposterior diffuse pleural thickening, priority diagnosis & 0.983 \\
x-ray chest anteroposterior distended bowel, priority diagnosis & 0.995 \\
x-ray chest anteroposterior electronic cardiac devices, priority diagnosis & 0.989 \\
x-ray chest anteroposterior focal airspace opacity, priority diagnosis & 0.914 \\
x-ray chest anteroposterior gallstones, priority diagnosis & 0.942 \\
x-ray chest anteroposterior gastric band, priority diagnosis & 0.964 \\
x-ray chest anteroposterior hiatal hernia, priority diagnosis & 0.994 \\
x-ray chest anteroposterior hyperinflation, priority diagnosis & 0.968 \\
x-ray chest anteroposterior image obscured, priority diagnosis & 0.991 \\
x-ray chest anteroposterior intercostal drain, priority diagnosis & 0.885 \\
x-ray chest anteroposterior intrathoracic foreign body, priority diagnosis & 0.795 \\
x-ray chest anteroposterior loculated effusion, priority diagnosis & 0.974 \\
x-ray chest anteroposterior lung collapse, priority diagnosis & 1.000 \\
x-ray chest anteroposterior mastectomy, priority diagnosis & 0.996 \\
x-ray chest anteroposterior mediastinal surgical clips, priority diagnosis & 0.945 \\
x-ray chest anteroposterior multifocal airspace opacities, priority diagnosis & 0.922 \\
x-ray chest anteroposterior multiple masses or nodules, priority diagnosis & 0.947 \\
x-ray chest anteroposterior optimal central venous catheter (cvc), priority diagnosis & 0.938 \\
x-ray chest anteroposterior optimal endotracheal tube (ett), priority diagnosis & 0.989 \\
x-ray chest anteroposterior optimal nasogastric tube (ngt), priority diagnosis & 0.971 \\
x-ray chest anteroposterior optimal pulmonary arterial catheter (pac), priority diagnosis & 0.984 \\
x-ray chest anteroposterior overexposed, priority diagnosis & 0.890 \\
x-ray chest anteroposterior patient rotation, priority diagnosis & 0.737 \\
x-ray chest anteroposterior pectus excavatum, priority diagnosis & 0.929 \\
x-ray chest anteroposterior peribronchial cuffing, priority diagnosis & 0.952 \\
x-ray chest anteroposterior perihilar airspace opacity, priority diagnosis & 0.996 \\

      \bottomrule
    \end{tabular}

  \end{threeparttable}
\end{table}

\begin{table}[h!]
  \centering
  \begin{threeparttable}
    \caption{\ourmodel MGB CXR Dataset Category Performance (2/2)}
    \label{tab:mgbcxrb}
    \begin{tabular}{@{}lc@{}}
      \toprule
      Category & AUC  \\
      \midrule 
x-ray chest anteroposterior pleural mass, priority diagnosis & 0.933 \\
x-ray chest anteroposterior pneumomediastinum, priority diagnosis & 0.979 \\
x-ray chest anteroposterior post resection volume loss, priority diagnosis & 0.998 \\
x-ray chest anteroposterior proper placement of pulmonary artery catheter, priority diagnosis & 0.992 \\
x-ray chest anteroposterior pulmonary artery enlargement, priority diagnosis & 0.965 \\
x-ray chest anteroposterior pulmonary congestion, priority diagnosis & 0.993 \\
x-ray chest anteroposterior rib fixation, priority diagnosis & 0.983 \\
x-ray chest anteroposterior rib lesion, priority diagnosis & 0.927 \\
x-ray chest anteroposterior rib resection, priority diagnosis & 0.985 \\
x-ray chest anteroposterior rotator cuff anchor, priority diagnosis & 0.961 \\
x-ray chest anteroposterior scapular fracture, priority diagnosis & 0.822 \\
x-ray chest anteroposterior scapular lesion, priority diagnosis & 0.311 \\
x-ray chest anteroposterior scoliosis, priority diagnosis & 0.967 \\
x-ray chest anteroposterior segmental collapse, priority diagnosis & 0.890 \\
x-ray chest anteroposterior shoulder arthritis, priority diagnosis & 0.899 \\
x-ray chest anteroposterior shoulder dislocation, priority diagnosis & 0.830 \\
x-ray chest anteroposterior shoulder fixation, priority diagnosis & 0.990 \\
x-ray chest anteroposterior shoulder replacement, priority diagnosis & 0.997 \\
x-ray chest anteroposterior simple effusion, priority diagnosis & 0.811 \\
x-ray chest anteroposterior simple pneumothorax, priority diagnosis & 0.945 \\
x-ray chest anteroposterior solitary lung mass, priority diagnosis & 0.929 \\
x-ray chest anteroposterior solitary lung nodule, priority diagnosis & 0.879 \\
x-ray chest anteroposterior solitary pulmonary nodule, priority diagnosis & 0.916 \\
x-ray chest anteroposterior spinal fixation, priority diagnosis & 0.997 \\
x-ray chest anteroposterior sternotomy wires, priority diagnosis & 0.993 \\
x-ray chest anteroposterior subcutaneous emphysema, priority diagnosis & 0.976 \\
x-ray chest anteroposterior subdiaphragmatic gas, priority diagnosis & 0.931 \\
x-ray chest anteroposterior suboptimal central venous catheter (cvc), priority diagnosis & 0.876 \\
x-ray chest anteroposterior suboptimal endotracheal tube (ett), priority diagnosis & 0.962 \\
x-ray chest anteroposterior suboptimal nasogastric tube (ngt), priority diagnosis & 0.912 \\
x-ray chest anteroposterior suboptimal pulmonary arterial catheter (pac), priority diagnosis & 0.983 \\
x-ray chest anteroposterior superior mediastinal mass, priority diagnosis & 0.776 \\
x-ray chest anteroposterior tension pneumothorax, priority diagnosis & 0.996 \\
x-ray chest anteroposterior tracheal deviation, priority diagnosis & 0.940 \\
x-ray chest anteroposterior underexposed, priority diagnosis & 0.989 \\
x-ray chest anteroposterior underinflation, priority diagnosis & 0.954 \\
x-ray chest anteroposterior unfolded aorta, priority diagnosis & 0.978 \\
x-ray chest anteroposterior upper zone bullae, priority diagnosis & 0.933 \\
x-ray chest anteroposterior upper zone fibrotic volume loss, priority diagnosis & 0.972 \\
x-ray chest anteroposterior widened aortic contour, priority diagnosis & 0.949 \\
x-ray chest anteroposterior widened cardiac silhouette, priority diagnosis & 0.970 \\

      \bottomrule
    \end{tabular}

  \end{threeparttable}
\end{table}

\begin{table}[h!]
  \centering
  \begin{threeparttable}
    \caption{\ourmodel RSNA Mammography \& BUSI Dataset Category Performance}
    \label{tab:rsnamamm}
    \begin{tabular}{@{}lc@{}}
      \toprule
      Category & AUC  \\
      \midrule 
x-ray breast (mamma) mammography benign with implant & 0.999 \\
x-ray breast (mamma) mammography benign without implant & 0.903 \\
x-ray breast (mamma) mammography cancer with implant & 0.989 \\
x-ray breast (mamma) mammography cancer without implant & 0.846 \\
ultrasound (US) breast benign & 0.967 \\
ultrasound (US) breast malignant & 0.985 \\
ultrasound (US) breast normal & 0.984 \\

      \bottomrule
    \end{tabular}

  \end{threeparttable}
\end{table}

\begin{table}[h!]
  \centering
  \begin{threeparttable}
    \caption{\ourmodel Exam Parameter Dataset Category Performance}
    \label{tab:examparam}
    \begin{tabular}{@{}lc@{}}
      \toprule
      Category & AUC  \\
      \midrule 
computed tomography abdomen axial, (CT ABDOMEN AXIAL) & 0.943 \\
computed tomography abdomen coronal, (CT ABDOMEN CORONAL) & 0.938 \\
computed tomography abdomen sagittal, (CT ABDOMEN SAGITTAL) & 0.990 \\
computed tomography brain axial, (CT BRAIN AXIAL) & 0.964 \\
computed tomography brain coronal, (CT BRAIN CORONAL) & 0.981 \\
computed tomography brain sagittal, (CT BRAIN SAGITTAL) & 0.962 \\
computed tomography chest axial, (CT CHEST AXIAL) & 0.965 \\
computed tomography chest coronal, (CT CHEST CORONAL) & 0.968 \\
computed tomography chest sagittal, (CT CHEST SAGITTAL) & 0.972 \\
computed tomography knee lower appendage axial, (CT KNEE AXIAL) & 0.974 \\
computed tomography knee lower appendage coronal, (CT KNEE CORONAL) & 0.993 \\
computed tomography knee lower appendage sagittal, (CT KNEE SAGITTAL) & 0.991 \\
magnetic resonance imaging abdomen axial, (MRI ABDOMEN AXIAL) & 0.981 \\
magnetic resonance imaging brain axial, (MRI BRAIN AXIAL) & 0.989 \\
magnetic resonance imaging brain coronal, (MRI BRAIN CORONAL) & 0.973 \\
magnetic resonance imaging brain sagittal, (MRI BRAIN SAGITTAL) & 0.986 \\
magnetic resonance imaging cspine sagittal, (MRI CSPINE SAGITTAL) & 0.992 \\
magnetic resonance imaging knee lower appendage axial, (MRI KNEE AXIAL) & 0.986 \\
magnetic resonance imaging knee lower appendage coronal, (MRI KNEE CORONAL) & 0.996 \\
magnetic resonance imaging knee lower appendage sagittal, (MRI KNEE SAGITTAL) & 0.981 \\
magnetic resonance imaging lspine lumbar spine sagittal, (MRI LSPINE SAGITTAL) & 0.992 \\

      \bottomrule
    \end{tabular}
  \end{threeparttable}
\end{table}

\begin{table}[h!]
  \centering
  \begin{threeparttable}
    \caption{\ourmodel HMC-QU Echo View Category Performance}
    \label{tab:hmcqu}
    \begin{tabular}{@{}lc@{}}
      \toprule
      Category & AUC  \\
      \midrule 
Ultrasound Echocardiography (UE) 2 Chamber (2CH) & 0.992 \\
Ultrasound Echocardiography (UE) 4 Chamber (4CH) & 0.981 \\

      \bottomrule
    \end{tabular}
  \end{threeparttable}
\end{table}

\begin{table}[h!]
  \centering
  \begin{threeparttable}
    \caption{\ourmodel Bing Echo View Category Performance}
    \label{tab:bingecho}
    \begin{tabular}{@{}lc@{}}
      \toprule
      Category & AUC  \\
      \midrule 
Ultrasound Echocardiography (UE) 2 Chamber (2CH) & 1.000 \\
Ultrasound Echocardiography (UE) 4 Chamber (4CH) & 0.832 \\
Ultrasound Echocardiography (UE) 5 Chamber (5CH) & 0.865 \\
Ultrasound Echocardiography (UE) CW-Doppler (CWD) & 0.995 \\
Ultrasound Echocardiography (UE) M-mode Doppler (MMD) & 0.992 \\
Ultrasound Echocardiography (UE) Parasternal Long Axis View (PLAX) & 0.943 \\
Ultrasound Echocardiography (UE) Short Axis (SA) & 0.961 \\

      \bottomrule
    \end{tabular}

  \end{threeparttable}
\end{table}

\begin{table}[h!]
  \centering
  \begin{threeparttable}
    \caption{\ourmodel ISIC2019 Category Performance}
    \label{tab:isic2019}
    \begin{tabular}{@{}lc@{}}
      \toprule
      Category & AUC  \\
      \midrule 
dermatology dermoscopy Actinic keratosis & 0.953 \\
dermatology dermoscopy Basal cell carcinoma & 0.985 \\
dermatology dermoscopy Benign keratosis & 0.952 \\
dermatology dermoscopy Dermatofibroma & 0.988 \\
dermatology dermoscopy Melanocytic nevus & 0.957 \\
dermatology dermoscopy Melanoma & 0.937 \\
dermatology dermoscopy Squamous cell carcinoma & 0.974 \\
dermatology dermoscopy Vascular lesion & 0.983 \\

      \bottomrule
    \end{tabular}

  \end{threeparttable}
\end{table}

\begin{table}[h!]
  \centering
  \begin{threeparttable}
    \caption{\ourmodel SD-198 Category Performance (1/4)}
    \label{tab:sd1981}
    \begin{tabular}{@{}lc@{}}
      \toprule
      Category & AUC  \\
      \midrule 

dermatology clinical photography Acne Keloidalis Nuchae & 0.999 \\
dermatology clinical photography Acne Vulgaris & 0.999 \\
dermatology clinical photography Acrokeratosis Verruciformis & 0.998 \\
dermatology clinical photography Actinic solar Damage(Actinic Cheilitis) & 1.000 \\
dermatology clinical photography Actinic solar Damage(Actinic Keratosis) & 0.984 \\
dermatology clinical photography Actinic solar Damage(Cutis Rhomboidalis Nuchae) & 1.000 \\
dermatology clinical photography Actinic solar Damage(Pigmentation) & 0.995 \\
dermatology clinical photography Actinic solar Damage(Solar Elastosis) & 0.999 \\
dermatology clinical photography Actinic solar Damage(Solar Purpura) & 0.998 \\
dermatology clinical photography Actinic solar Damage(Telangiectasia) & 0.998 \\
dermatology clinical photography Acute Eczema & 0.999 \\
dermatology clinical photography Allergic Contact Dermatitis & 0.928 \\
dermatology clinical photography Alopecia Areata & 1.000 \\
dermatology clinical photography Androgenetic Alopecia & 1.000 \\
dermatology clinical photography Angioma & 0.978 \\
dermatology clinical photography Angular Cheilitis & 0.995 \\
dermatology clinical photography Aphthous Ulcer & 1.000 \\
dermatology clinical photography Apocrine Hydrocystoma & 0.997 \\
dermatology clinical photography Arsenical Keratosis & 0.988 \\
dermatology clinical photography Balanitis Xerotica Obliterans & 1.000 \\
dermatology clinical photography Basal Cell Carcinoma & 0.981 \\
dermatology clinical photography Beau's Lines & 0.997 \\
dermatology clinical photography Becker's Nevus & 0.999 \\
dermatology clinical photography Behcet's Syndrome & 0.999 \\
dermatology clinical photography Benign Keratosis & 0.976 \\
dermatology clinical photography Blue Nevus & 0.985 \\
dermatology clinical photography Bowenoid Papulosis & 1.000 \\
dermatology clinical photography Bowen's Disease & 0.981 \\
dermatology clinical photography Cafe Au Lait Macule & 1.000 \\
dermatology clinical photography Callus & 0.999 \\
dermatology clinical photography Candidiasis & 0.994 \\
dermatology clinical photography Cellulitis & 0.996 \\
dermatology clinical photography Chalazion & 1.000 \\
dermatology clinical photography Clubbing of Fingers & 0.992 \\
dermatology clinical photography Compound Nevus & 0.990 \\
dermatology clinical photography Congenital Nevus & 0.999 \\
dermatology clinical photography Crowe's Sign & 1.000 \\
dermatology clinical photography Cutanea Larva Migrans & 1.000 \\
dermatology clinical photography Cutaneous Horn & 0.998 \\
dermatology clinical photography Cutaneous T-Cell Lymphoma & 0.992 \\
dermatology clinical photography Cutis Marmorata & 0.997 \\
dermatology clinical photography Darier-White Disease & 0.973 \\
dermatology clinical photography Dermatofibroma & 0.995 \\
dermatology clinical photography Dermatosis Papulosa Nigra & 1.000 \\
dermatology clinical photography Desquamation & 1.000 \\
dermatology clinical photography Digital Fibroma & 0.996 \\
dermatology clinical photography Dilated Pore of Winer & 0.972 \\
dermatology clinical photography Discoid Lupus Erythematosus & 1.000 \\
dermatology clinical photography Disseminated Actinic Porokeratosis & 1.000 \\
dermatology clinical photography Drug Eruption & 0.968 \\
      \bottomrule
    \end{tabular}
  \end{threeparttable}
\end{table}

\begin{table}[h!]
  \centering
  \begin{threeparttable}
    \caption{\ourmodel SD-198 Category Performance (2/4)}
    \label{tab:sd1982}
    \begin{tabular}{@{}lc@{}}
      \toprule
      Category & AUC  \\
      \midrule 
dermatology clinical photography Drug Eruption & 0.968 \\
dermatology clinical photography Dry Skin Eczema & 0.979 \\
dermatology clinical photography Dyshidrosiform Eczema & 0.995 \\
dermatology clinical photography Dysplastic Nevus & 0.992 \\
dermatology clinical photography Eccrine Poroma & 0.996 \\
dermatology clinical photography Eczema & 0.983 \\
dermatology clinical photography Epidermal Nevus & 0.987 \\
dermatology clinical photography Epidermoid Cyst & 0.992 \\
dermatology clinical photography Epithelioma Adenoides Cysticum & 0.987 \\
dermatology clinical photography Erythema Ab Igne & 0.921 \\
dermatology clinical photography Erythema Annulare Centrifigum & 0.973 \\
dermatology clinical photography Erythema Craquele & 0.987 \\
dermatology clinical photography Erythema Multiforme & 0.969 \\
dermatology clinical photography Exfoliative Erythroderma & 0.993 \\
dermatology clinical photography Factitial Dermatitis & 0.994 \\
dermatology clinical photography Favre Racouchot & 0.999 \\
dermatology clinical photography Fibroma & 1.000 \\
dermatology clinical photography Fibroma Molle & 0.999 \\
dermatology clinical photography Fixed Drug Eruption & 0.924 \\
dermatology clinical photography Follicular Mucinosis & 0.956 \\
dermatology clinical photography Follicular Retention Cyst & 0.985 \\
dermatology clinical photography Fordyce Spots & 1.000 \\
dermatology clinical photography Frictional Lichenoid Dermatitis & 0.999 \\
dermatology clinical photography Ganglion & 0.999 \\
dermatology clinical photography Geographic Tongue & 1.000 \\
dermatology clinical photography Granulation Tissue & 0.985 \\
dermatology clinical photography Granuloma Annulare & 0.984 \\
dermatology clinical photography Green Nail & 0.995 \\
dermatology clinical photography Guttate Psoriasis & 0.997 \\
dermatology clinical photography Hailey Hailey Disease & 0.969 \\
dermatology clinical photography Half and Half Nail & 0.998 \\
dermatology clinical photography Halo Nevus & 0.994 \\
dermatology clinical photography Herpes Simplex Virus & 0.977 \\
dermatology clinical photography Herpes Zoster & 0.994 \\
dermatology clinical photography Hidradenitis Suppurativa & 0.999 \\
dermatology clinical photography Histiocytosis X & 1.000 \\
dermatology clinical photography Hyperkeratosis Palmaris Et Plantaris & 0.998 \\
dermatology clinical photography Hypertrichosis & 0.986 \\
dermatology clinical photography Ichthyosis & 0.998 \\
dermatology clinical photography Impetigo & 0.985 \\
dermatology clinical photography Infantile Atopic Dermatitis & 0.994 \\
dermatology clinical photography Inverse Psoriasis & 0.978 \\
dermatology clinical photography Junction Nevus & 0.933 \\
dermatology clinical photography Keloid & 0.996 \\
dermatology clinical photography Keratoacanthoma & 0.975 \\
dermatology clinical photography Keratolysis Exfoliativa of Wende & 1.000 \\
dermatology clinical photography Keratosis Pilaris & 0.998 \\
dermatology clinical photography Kerion & 0.998 \\
dermatology clinical photography Koilonychia & 0.997 \\
dermatology clinical photography Kyrle's Disease & 1.000 \\
dermatology clinical photography Leiomyoma & 1.000 \\

      \bottomrule
    \end{tabular}
  \end{threeparttable}
\end{table}

\begin{table}[h!]
  \centering
  \begin{threeparttable}
    \caption{\ourmodel SD-198 Category Performance (3/4)}
    \label{tab:sd1983}
    \begin{tabular}{@{}lc@{}}
      \toprule
      Category & AUC  \\
      \midrule 

dermatology clinical photography Lentigo Maligna Melanoma & 0.994 \\
dermatology clinical photography Leukocytoclastic Vasculitis & 0.977 \\
dermatology clinical photography Leukonychia & 0.997 \\
dermatology clinical photography Lichen Planus & 0.900 \\
dermatology clinical photography Lichen Sclerosis Et Atrophicus & 0.986 \\
dermatology clinical photography Lichen Simplex Chronicus & 0.950 \\
dermatology clinical photography Lichen Spinulosis & 1.000 \\
dermatology clinical photography Linear Epidermal Nevus & 0.991 \\
dermatology clinical photography Lipoma & 0.999 \\
dermatology clinical photography Livedo Reticularis & 1.000 \\
dermatology clinical photography Lymphangioma Circumscriptum & 1.000 \\
dermatology clinical photography Lymphocytic Infiltrate of Jessner & 0.994 \\
dermatology clinical photography Lymphomatoid Papulosis & 0.999 \\
dermatology clinical photography Malignant Melanoma & 0.990 \\
dermatology clinical photography Mal Perforans & 1.000 \\
dermatology clinical photography Median Nail Dystrophy & 0.999 \\
dermatology clinical photography Melasma & 0.998 \\
dermatology clinical photography Metastatic Carcinoma & 0.980 \\
dermatology clinical photography Milia & 0.960 \\
dermatology clinical photography Molluscum Contagiosum & 0.971 \\
dermatology clinical photography Morphea & 0.942 \\
dermatology clinical photography Mucha Habermann Disease & 0.931 \\
dermatology clinical photography Mucous Membrane Psoriasis & 0.998 \\
dermatology clinical photography Myxoid Cyst & 0.998 \\
dermatology clinical photography Nail Dystrophy & 0.993 \\
dermatology clinical photography Nail Nevus & 1.000 \\
dermatology clinical photography Nail Psoriasis & 0.986 \\
dermatology clinical photography Nail Ridging & 0.996 \\
dermatology clinical photography Neurodermatitis & 0.949 \\
dermatology clinical photography Neurofibroma & 0.986 \\
dermatology clinical photography Neurotic Excoriations & 0.996 \\
dermatology clinical photography Nevus Comedonicus & 0.973 \\
dermatology clinical photography Nevus Incipiens & 0.981 \\
dermatology clinical photography Nevus Sebaceous of Jadassohn & 0.992 \\
dermatology clinical photography Nevus Spilus & 1.000 \\
dermatology clinical photography Nummular Eczema & 0.991 \\
dermatology clinical photography Onychogryphosis & 1.000 \\
dermatology clinical photography Onycholysis & 0.998 \\
dermatology clinical photography Onychomycosis & 0.999 \\
dermatology clinical photography Onychoschizia & 0.999 \\
dermatology clinical photography Paronychia & 0.995 \\
dermatology clinical photography Pearl Penile Papules & 1.000 \\
dermatology clinical photography Perioral Dermatitis & 0.997 \\
dermatology clinical photography Pincer Nail Syndrome & 0.993 \\
dermatology clinical photography Pitted Keratolysis & 1.000 \\
dermatology clinical photography Pityriasis Alba & 1.000 \\
dermatology clinical photography Pityriasis Rosea & 0.976 \\
dermatology clinical photography Pityrosporum Folliculitis & 0.990 \\
dermatology clinical photography Poikiloderma Atrophicans Vasculare & 0.996 \\
dermatology clinical photography Pomade Acne & 1.000 \\

      \bottomrule
    \end{tabular}
  \end{threeparttable}
\end{table}

\begin{table}[h!]
  \centering
  \begin{threeparttable}
    \caption{\ourmodel SD-198 Category Performance (4/4)}
    \label{tab:sd1984}
    \begin{tabular}{@{}lc@{}}
      \toprule
      Category & AUC  \\
      \midrule 
dermatology clinical photography Pseudofolliculitis Barbae & 0.998 \\
dermatology clinical photography Pseudorhinophyma & 0.998 \\
dermatology clinical photography Psoriasis & 0.987 \\
dermatology clinical photography Pustular Psoriasis & 0.987 \\
dermatology clinical photography Pyoderma Gangrenosum & 0.959 \\
dermatology clinical photography Pyogenic Granuloma & 0.996 \\
dermatology clinical photography Racquet Nail & 0.990 \\
dermatology clinical photography Radiodermatitis & 0.992 \\
dermatology clinical photography Rhinophyma & 0.997 \\
dermatology clinical photography Rosacea & 0.990 \\
dermatology clinical photography Scalp Psoriasis & 0.999 \\
dermatology clinical photography Scar & 0.984 \\
dermatology clinical photography Scarring Alopecia & 0.994 \\
dermatology clinical photography Schamberg's Disease & 0.991 \\
dermatology clinical photography Sebaceous Gland Hyperplasia & 0.994 \\
dermatology clinical photography Seborrheic Dermatitis & 0.964 \\
dermatology clinical photography Seborrheic Keratosis & 0.988 \\
dermatology clinical photography Skin Tag & 0.998 \\
dermatology clinical photography Solar Lentigo & 0.992 \\
dermatology clinical photography Stasis Dermatitis & 0.990 \\
dermatology clinical photography Stasis Edema & 0.992 \\
dermatology clinical photography Stasis Ulcer & 0.998 \\
dermatology clinical photography Steroid Acne & 0.999 \\
dermatology clinical photography Steroid Striae & 0.998 \\
dermatology clinical photography Steroid Use abusemisuse Dermatitis & 0.991 \\
dermatology clinical photography Stomatitis & 1.000 \\
dermatology clinical photography Strawberry Hemangioma & 0.999 \\
dermatology clinical photography Striae & 1.000 \\
dermatology clinical photography Subungual Hematoma & 1.000 \\
dermatology clinical photography Syringoma & 1.000 \\
dermatology clinical photography Terry's Nails & 1.000 \\
dermatology clinical photography Tinea Corporis & 0.992 \\
dermatology clinical photography Tinea Cruris & 1.000 \\
dermatology clinical photography Tinea Faciale & 0.992 \\
dermatology clinical photography Tinea Manus & 0.997 \\
dermatology clinical photography Tinea Pedis & 0.993 \\
dermatology clinical photography Tinea Versicolor & 1.000 \\
dermatology clinical photography Toe Deformity & 0.999 \\
dermatology clinical photography Trichilemmal Cyst & 1.000 \\
dermatology clinical photography Trichofolliculoma & 0.997 \\
dermatology clinical photography Trichostasis Spinulosa & 1.000 \\
dermatology clinical photography Ulcer & 0.996 \\
dermatology clinical photography Urticaria & 0.935 \\
dermatology clinical photography Varicella & 0.976 \\
dermatology clinical photography Verruca Vulgaris & 0.908 \\
dermatology clinical photography Vitiligo & 0.970 \\
dermatology clinical photography Wound Infection & 0.999 \\
dermatology clinical photography Xerosis & 0.999 \\

      \bottomrule
    \end{tabular}
  \end{threeparttable}
\end{table}

\begin{table}[h!]
  \centering
  \begin{threeparttable}
    \caption{\ourmodel PADUFES20 Category Performance}
    \label{tab:padufes20}
    \begin{tabular}{@{}lc@{}}
      \toprule
      Category & AUC  \\
      \midrule 
dermatology clinical photography Actinic Keratosis (ACK) cropped & 0.936 \\
dermatology clinical photography Basal Cell Carcinoma (BCC) cropped & 0.939 \\
dermatology clinical photography Malignant Melanoma (MEL) cropped & 0.993 \\
dermatology clinical photography Nevus (NEV) cropped & 0.986 \\
dermatology clinical photography Seborrheic Keratosis (SEK) cropped & 0.980 \\
dermatology clinical photography Squamous Cell Carcinoma (SCC) cropped & 0.888 \\

      \bottomrule
    \end{tabular}
  \end{threeparttable}
\end{table}

\begin{table}[h!]
  \centering
  \begin{threeparttable}
    \caption{\ourmodel PADUFES20 (No Patient Overlap*) Category Performance}
    \label{tab:padufes20npo}
    \begin{tabular}{@{}lc@{}}
      \toprule
      Category & AUC  \\
      \midrule 
dermatology clinical photography Actinic Keratosis (ACK) cropped & 0.976 \\
dermatology clinical photography Basal Cell Carcinoma (BCC) cropped & 0.979 \\
dermatology clinical photography Malignant Melanoma (MEL) cropped & 0.991 \\
dermatology clinical photography Nevus (NEV) cropped & 0.975 \\
dermatology clinical photography Seborrheic Keratosis (SEK) cropped & 0.948 \\
dermatology clinical photography Squamous Cell Carcinoma (SCC) cropped & 0.937 \\

      \bottomrule
    \end{tabular}
    \begin{tablenotes}
      \small
      \item[*] Test set stripped of any patients present in training. 
    \end{tablenotes}
  \end{threeparttable}
\end{table}

\begin{table}[h!]
  \centering
  \begin{threeparttable}
    \caption{\ourmodel OCT2017 Category Performance}
    \label{tab:oct2017}
    \begin{tabular}{@{}lc@{}}
      \toprule
      Category & AUC  \\
      \midrule 

Optical Coherence Tomography (OCT), Retina, choroidal neovascularization (CNV) & 1.000 \\
Optical Coherence Tomography (OCT), Retina, diabetic macular edema (DME) & 1.000 \\
Optical Coherence Tomography (OCT), Retina, drusen (DRUSEN) & 1.000 \\
Optical Coherence Tomography (OCT), Retina, normal retina (NORMAL) & 1.000 \\

      \bottomrule
    \end{tabular}
  \end{threeparttable}
\end{table}

\begin{table}[h!]
  \centering
  \begin{threeparttable}
    \caption{\ourmodel OCT2018 Category Performance}
    \label{tab:oct2018}
    \begin{tabular}{@{}lc@{}}
      \toprule
      Category & AUC  \\
      \midrule 

Optical Coherence Tomography (OCT), Retina, choroidal neovascularization (CNV) & 0.997 \\
Optical Coherence Tomography (OCT), Retina, diabetic macular edema (DME) & 1.000 \\
Optical Coherence Tomography (OCT), Retina, drusen (DRUSEN) & 0.999 \\
Optical Coherence Tomography (OCT), Retina, normal retina (NORMAL) & 1.000 \\

      \bottomrule
    \end{tabular}
  \end{threeparttable}
\end{table}

\begin{table}[h!]
  \centering
  \begin{threeparttable}
    \caption{\ourmodel ODIR5K Category* Performance}
    \label{tab:odir5k}
    \begin{tabular}{@{}lc@{}}
      \toprule
      Category & AUC  \\
      \midrule 

retinal fundus branch retinal vein occlusion & 1.000 \\
retinal fundus cataract & 0.995 \\
retinal fundus chorioretinal atrophy & 0.994 \\
retinal fundus diabetic retinopathy & 0.807 \\
retinal fundus drusen & 0.915 \\
retinal fundus dry age-related macular degeneration & 0.931 \\
retinal fundus epiretinal membrane & 0.966 \\
retinal fundus epiretinal membrane over the macula & 0.993 \\
retinal fundus glaucoma & 0.955 \\
retinal fundus hypertensive retinopathy & 0.880 \\
retinal fundus low image quality & 0.960 \\
retinal fundus macular epiretinal membrane & 0.985 \\
retinal fundus maculopathy & 0.884 \\
retinal fundus mild nonproliferative retinopathy & 0.851 \\
retinal fundus moderate non proliferative retinopathy & 0.945 \\
retinal fundus myelinated nerve fibers & 0.997 \\
retinal fundus normal fundus & 0.889 \\
retinal fundus pathological myopia & 0.999 \\
retinal fundus proliferative diabetic retinopathy & 0.989 \\
retinal fundus refractive media opacity & 0.968 \\
retinal fundus retinal pigmentation & 0.804 \\
retinal fundus retinitis pigmentosa & 0.994 \\
retinal fundus severe nonproliferative retinopathy & 0.979 \\
retinal fundus spotted membranous change & 0.963 \\
retinal fundus suspected glaucoma & 0.978 \\
retinal fundus tessellated fundus & 0.988 \\
retinal fundus vitreous degeneration & 0.995 \\
retinal fundus wet age-related macular degeneration & 0.984 \\

      \bottomrule
    \end{tabular}
    \begin{tablenotes}
      \small
      \item[*] 79 categories trained, but 28 evaluated with more than 10 samples. 
    \end{tablenotes}
  \end{threeparttable}
\end{table}

\end{document}